\documentclass[
reprint,
superscriptaddress,
 amsmath,amssymb,
 aps,
]{revtex4-2}

\usepackage{graphicx}
\usepackage{dcolumn}
\usepackage{bm}
\usepackage{hyperref}

\begin{document}

\preprint{APS/123-QED}

\title{Direct Determination of Photonic Stopband Topological Character: A Framework based on Dispersion Measurements}

\author{Nitish Kumar Gupta}
\email{nitishkg@iisc.ac.in}
\affiliation{Centre for Lasers \& Photonics, Indian Institute of Technology Kanpur, 208016, India}

\author{Sapireddy Srinivasu}
\affiliation{Centre for Lasers \& Photonics, Indian Institute of Technology Kanpur, 208016, India}

\author{Mukesh Kumar}
\affiliation{CSIR-Central Scientific Instruments Organisation, Chandigarh, 160030, India}

\author{Anjani Kumar Tiwari}
\affiliation{Department of Physics, Indian Institute of Technology Roorkee, 247667, India}

\author{Sudipta Sarkar Pal}
\affiliation{CSIR-Central Scientific Instruments Organisation, Chandigarh, 160030, India} 

\author{Harshawardhan Wanare}
\affiliation{Centre for Lasers \& Photonics, Indian Institute of Technology Kanpur, 208016, India}
\affiliation{Department of Physics, Indian Institute of Technology Kanpur, 208016, India}

\author{S. Anantha Ramakrishna}
\affiliation{CSIR-Central Scientific Instruments Organisation, Chandigarh, 160030, India}
\affiliation{Department of Physics, Indian Institute of Technology Kanpur, 208016, India}

\date{\today}

\begin{abstract}
Ascertainment of photonic stopband absolute topological character requires information regarding the Bloch eigenfunction spatial distribution. Consequently, the experimental investigations predominantly restrict themselves to the bulk-boundary correspondence principle and the ensuing emergence of topological surface state. Although capable of establishing the equivalence/inequivalence of bandgaps, the determination of their absolute topological identity remains out of its purview. The alternate method of reflection phase-based identification also provides only contentious improvements owing to the measurement complexities pertaining to the interferometric setups. To circumvent these limitations, we resort to the Kramers-Kronig amplitude-phase causality considerations and propose an experimentally conducive method for bandgap topological character determination directly from the parametric reflectance measurements. Particularly, it has been demonstrated that in case of one-dimensional photonic crystals, polarization-resolved dispersion measurements suffice in qualitatively determining bandgaps’ absolute topological identities. By invoking the translational invariance of the investigated samples, we also define a parameter “differential effective mass,” that encapsulates bandgaps’ topological identities and engenders an experimentally discernible bandgap classifier. 

\end{abstract}
\maketitle

\section{Introduction}

Photonic topological insulators (PTIs) constitute a prolific platform to study the bulk band geometric phase aspects and the accompanying fundamental novel phenomena~\cite{kim2020recent,cheng2022asymmetric,tang2022topological,yang2017direct,ke2016topological,chen2019distinguishing,cohen2020generalized,xue2021topological}. Apart from this, they are also a promising candidate in applications like robust waveguiding~\cite{yang2020terahertz}, quantum information processing~\cite{nemoto2014photonic,chen2021topologically}, and volitional realization of edge/surface states~\cite{st2017lasing,xiao2014surface,hu2021double,tan2021topological,lin2021conjugated,hu2019strong,gupta2022realizing}. In recent years, PTIs have been realized with a variety of structural templates, including electromagnetic composites such as photonic crystals (PhCs)  $\&$ metamaterials (MMs)~\cite{khanikaev2013photonic,yang2019realization,lustig2019photonic,kim2022three,tan2022interfacial,liu2022topological,chen2014experimental,kumar2022topological}, and engineered artificial lattices~\cite{kurganov2022temperature,yang2020photonic,pocock2018topological,st2021measuring,rappoport2021topological,liu2021frequency,xia2020nontrivial}. In all these cases, the non-trivial band topology of PTIs and the resulting topological properties of bandgaps are characterized by quantized topological invariants whose determination constitutes a crucial step in understanding the attributes of PTIs. For example, in one of their simplest rendition as one-dimensional (1D) topological photonic crystal, the geometric Zak phase~\cite{zak1989berry,atala2013direct} is recognized as the relevant topological invariant that can be determined by keeping track of singularities of Bloch eigenfunction across the Brillouin zone. The topological character of associated stopbands, in this case, can be theoretically ascertained by tracking the evolution of cumulative Zak phases~\cite{xiao2014surface}. However, not being an observable of the system, an experimental effort to determine the topological invariants and bandgap implicit topological character has to invariably rely on indirect measurements where eigenfunction topology leaves its imprints~\cite{atala2013direct,jiao2021experimentally,abanin2013interferometric,mittal2016measurement,duca2015aharonov,flurin2017observing,ramasesh2017direct,von2021measurement,zhang2017direct,hu2015measurement,hafezi2014measuring,wang2019direct,cardano2017detection,grusdt2014measuring,shen2018topological}. Specifically, in 1D PhCs, determination of stopband topological identity requires accessing the dynamic phase response associated to the scattering parameters~\cite{wang2016measurement,wang2017optical,gao2015determination,gupta2022spectroscopic}. This necessitates interferometric setups and a reliance on relative measurements. Although some of the complexities can be relieved by resorting to the bulk-boundary correspondence principle where interferometric setups are not required, and atleast the equivalence/inequivalence of bandgaps' topological characters can be stated from reflectance measurement alone (by locating a midgap state), its experimental implementation in bosonic systems is inexpedient as it requires concatenation of bandgaps.

\noindent In a significant departure, we develop a neoteric methodology to determine the bandgap topological character of 1D PhCs directly from the dispersion diagram and experimentally demonstrate it at optical frequencies. Our proposition completely abrogates the requirement of phase information or Bloch eigenfunction distribution (and hence, interferometric measurements). Specifically, we resort to the polarization-resolved angular dispersion measurements that provide an experimentally amenable and unambiguous signature for bandgap topological character identification. The genesis of our proposal lies in Kramers-Kronig causality considerations~\cite{toll1956causality,ramakrishna2008physics}, which establish a formal interdependence between the amplitude and phase response, permitting us to extract sufficient information regarding the eigenfunction topology from reflectance (intensity) measurements itself. Discriminatory contributions for topologically distinct bandgaps arise on account of their distinct phase dispersions. Notably, we observe the presence of topological defects in the form of wrapping cut discontinuities (WCDs)~\cite{venema2008optical} in the parametric reflection phase plots, altogether leaving a measurable impact on the reflection intensity. Although feeble at near-normal incidence, the impressions become prominent at higher incidence angles which have been captured by polarization-resolved angular dispersion measurements. Crucially, unlike other techniques, our method can not only make out a distinction in the topological character of bandgaps, but it provides enough information to ascertain the absolute topological identity of a bandgap. We have experimentally verified these propositions and based on their peculiar dispersion labeled them ``ENG-like bandgaps" (exhibiting electric conductor-like response) and ``MNG-like bandgaps" (exhibiting magnetic conductor-like response). Finally, by exploiting the fact that the translational invariance in planar samples upholds the conservation of in-plane component of photon momentum, we define a ``differential effective mass parameter" for the bandgaps that captures the implicit topological identities of bandgaps in direct commensuration with Dirac equation formalism and provides us with a new classifier for topological order characterization.

In order to fathom the repercussions of bandgap topological character on measurable system parameters, it is crucial to identify an appropriate parameter space that harbingers traceable signatures. This task has been undertaken in the next section where we commence with calculations corresponding to bulk structures and later on uncover some crucial connections between bulk behaviour and surface properties.

\section{Results and Discussion}

\subsection{Theoretical assessment of Topological Order: An alternate photonic bandstructure and phase transition phenomenon}\label{sec1_1}

In condensed matter systems, we often employ the Dirac Hamiltonian to describe the properties of topologically non-trivial insulators. The effective mass appearing in this description permits us to capture the distinction in the spin-1/2 particle/antiparticle wavefunctions by means of a change in sign. Specifically, in one-dimensional (1D) configuration, the Dirac equation refers to an eigenvalue problem of the form~\cite{asboth2016short}:
\begin{equation}
     \left(-i \hbar \sigma_{x}v\frac{\partial}{\partial x} + m_{eff}(x)v^{2}\sigma_{z} + V(x) \right)\mid \psi \rangle = i \hbar \frac{\partial }{\partial t}\mid \psi \rangle
     \label{eq1}
\end{equation}

\begin{figure}[h]%
\centering
\includegraphics[width=0.5\textwidth]{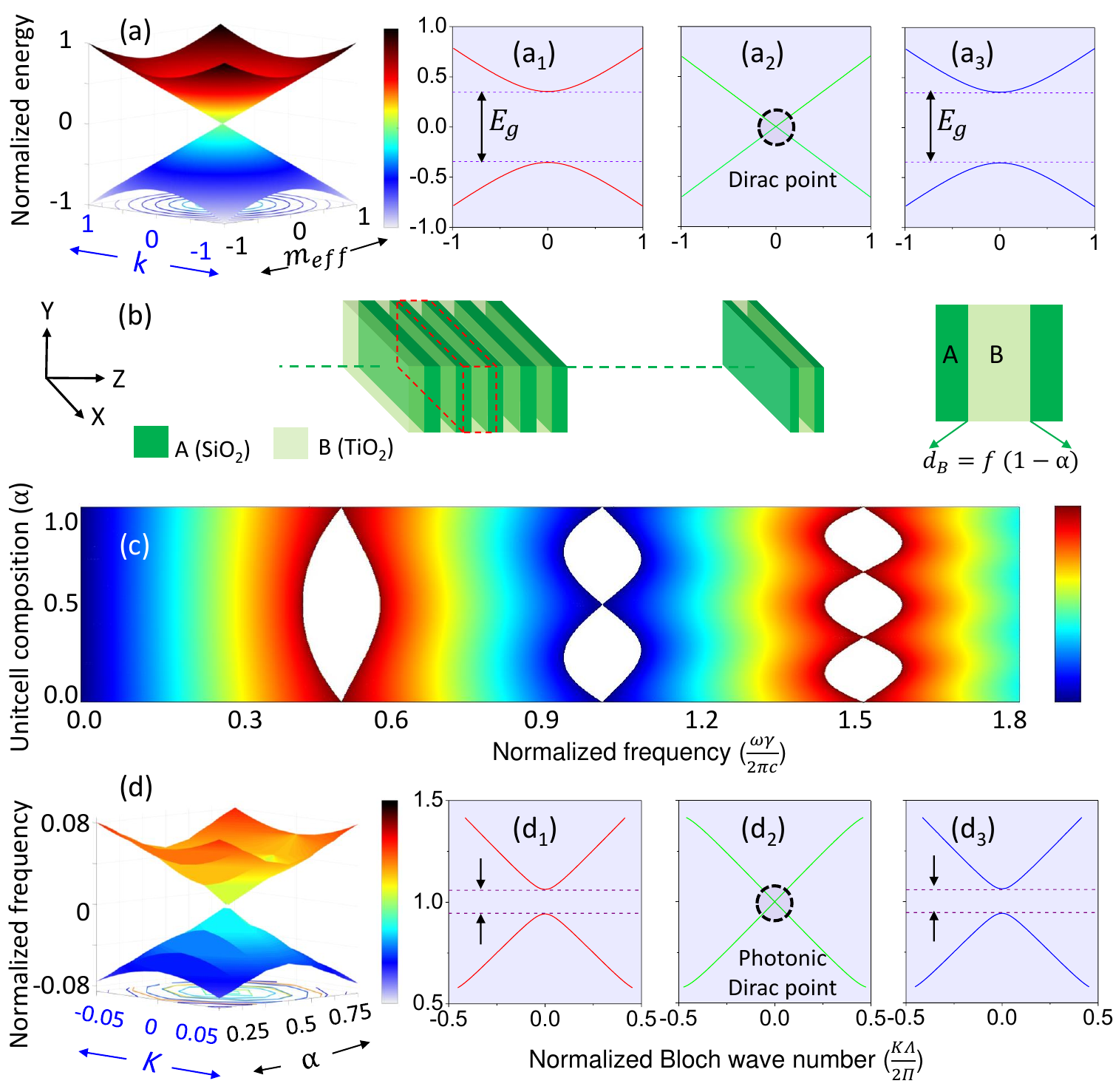}
\caption{(a) Dirac equation eigenvalue spectrum in ($m_{eff}-k$) 2D parameter space; $(a_1),~(a_2)~\&~(a_3)$ depict the band diagrams for $m_{eff}=-0.50,~0.00$ and $0.50$, respectively. (b) A representative schematic of the PhC structure with the axes convention. (c) PhC eigenvalue spectrum in unitcell fractional composition-normalized frequency ($\alpha-\omega_N$) 2D parameter space. (d) PhC eigenvalue spectrum in conventional settings: it exhibits a Dirac-like dispersion in ($\alpha-K$) 2D parameter space in the neighborhood of $\alpha=0.50$; $(d_1),~(d_2)~\&~(d_3)$ depict the band diagrams for $\alpha=0.25,~0.50$ and  $0.75$, respectively.}
\label{fig1}
\end{figure} 

\noindent where the parameter ${m_{eff}}(x)$ is recognized as the effective mass that can take positive and negative values for particles and antiparticles. Under certain simplifying assumptions the energy eigenvalues for the particle state turn out to be  $E_{1,2}=\pm \sqrt{m^{2}_{eff} + p^{2}}$ with the corresponding stationary eigenstates being $\mid \psi \rangle_{1,2} =\begin{pmatrix} 
    \frac{-p}{m_{eff} \pm \sqrt{m^{2}_{eff} + p^{2}}} ~~1
    \end{pmatrix}^T$. 
 From these expressions, we notice that the energy eigenvalues, which define the observables of the system, are incapable of perceiving any sign change in ${m_{eff}}(x)$. Only when we resort to the eigenstates ($\mid \psi \rangle$) does a change of sign leaves an impression. To succinctly highlight this point, in Fig.~\ref{fig1}(a), we plot the energy eigenvalues (${E}$) in the ${m_{eff}-k}$ two-dimensional (2D) parameter space with ${m_{eff}}$ varying from $-1$ to $+1$ in normalized units. As evident from the plots, the energy eigenvalues remain indiscernible as ${m_{eff}}$ transits from positive to negative values across the ${m_{eff}=0}$ point. Besides, the plots also indicate that a sign change in ${m_{eff}}$ accompanies the transition through a bandgap closing point (known as the Dirac point). It is explicitly highlighted in Fig.~\ref{fig1}$(a_1)-(a_3)$ by plotting the ${E-k}$ band diagrams for ${m_{eff}}=-0.50,~0.00$ and $0.50$ respectively. Therefore, we notice that the Dirac point (${m_{eff}}=0$) in the band diagram serves as a site for a (topological) phase transition for particle wavefunction, explaining its indispensable role in topological studies. 
 
Search for topologically non-trivial behaviour in PhCs usually take advantage of this precedence where we begin by locating a bandgap closing point in the photonic bandstructure (a potential site for topological phase transition) and thereafter resort to topological band theory for confirmations and further investigations. To obtain bandgap closing points in a deterministic manner, we define a parameter $\alpha=\frac{n_{A}d_{A}}{n_{A}d_{A}+n_{B}d_{B}}=\frac{n_{A}d_{A}}{\gamma}$ for the PhC. We call it the ``unit cell fractional composition parameter" and calculate the PhC dispersion in terms of it. Using Eq. 13 of Supporting Information file, in Fig.~\ref{fig1}(c), we plot the bandstructure for the 1D PhC in a new $\alpha-\omega_N$ 2D parameter space ($\omega_N$ being the normalized frequency). This non-routine bandstructure has the merit of explicitly bringing out the bandgap closing points for all the bandgaps. For example, the second-order bandgap exhibits gap closing at $\alpha=0.50$ while the third-order bandgap exhibits the gap closing at $\alpha=0.33$ and $\alpha=0.66$. We call these bandgap closing points as nodal points. A closer look at the Fig.~\ref{fig1}(c) bandstructure allows us to also propose an empirical relation for the normalized bandgap width ($\delta\omega_N$) in terms of $\alpha$ as $\delta\omega_N(\alpha)= A\sin{n\pi\alpha}$ ($n$ denoting the bandgap number). From this relation, we see that for any nth order bandgap, the nodal points divide the $\alpha$ expanse in equal proportions. For example, we observe in Fig.~\ref{fig1}(c) that the nodal point for the second-order bandgap leads to a mirror-symmetric bandgap splitting around the $\alpha=0.50$ line. This reminds us the symmetric nature of Dirac dispersion in the vicinity of ${m_{eff}=0}$, encouraging us to call the nodal point at $\alpha=0.50$ (in Fig.~\ref{fig1}(c)) as a photonic Dirac point. To further concretize this correspondence, in Eq. 18 of Supporting Information, we  establish that in the vicinity of nodal point $\alpha=0.50$, the PhC dispersion relation can also be reduced to a linear form. This expression promulgates a Dirac-like linear band crossing for photonic bands, as evident in Fig.~\ref{fig1}(d). In more details, Fig.~\ref{fig1}$(d_1)-(d_3)$ explicitly depict ${\omega_N-K}$ band diagrams for three specific cases corresponding to ${\alpha}= 0.25, 0.50$ and $0.75$ respectively. A direct comparison between Fig.~\ref{fig1}$(a)-(a_3)$ and Fig.~\ref{fig1}$(d)-(d_3)$ suggests that the role exhibited by the effective mass $m_{eff}$ in the case of Dirac particles has been taken over by the parameter $\alpha$ in case of photon propagation in PhCs. Therefore, we predict that the nodal points obtained from the bandstructure of Fig.~\ref{fig1}(c) form a potential site for topological phase transition. As a next step, we rigorously examine this prediction by evolving the bandstructure across the $\alpha=0.50$ nodal point and track the evolution of bulk-band geometric Zak phases (topological invariant in 1D system), calculated using the expression provided in Section IV of Supporting Information file.

\begin{figure}[h]%
\centering
\includegraphics[width=0.5\textwidth]{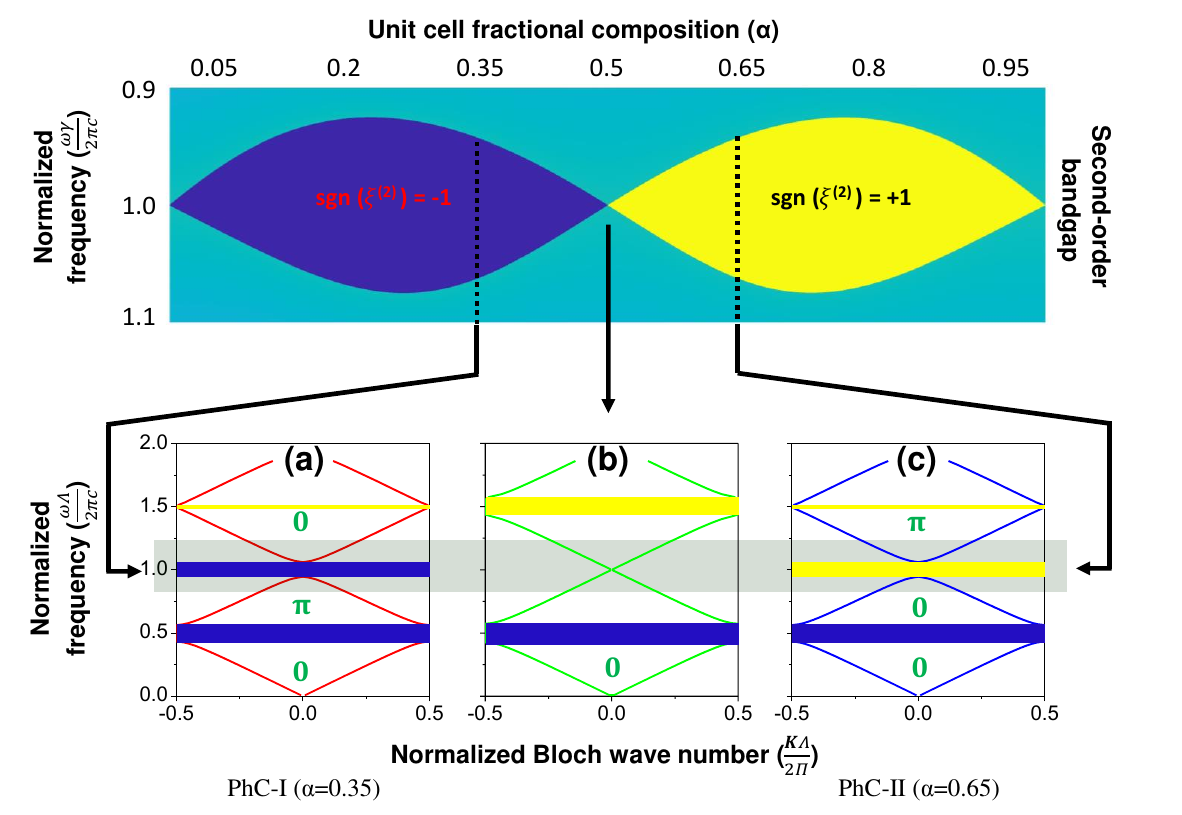}
\caption{Theoretical determination of topological identities of bandgaps for the bulk PhC by tracking cumulative accumulation of the Zak phases for (a) $\alpha=0.35$, (b) $\alpha=0.50$ and (c) $\alpha=0.65$, respectively (the two complementary bandgaps have been depicted with binary color codes consisting of blue and yellow colors). }
\label{fig2}
\end{figure}

\noindent The calculation results are provided in Fig.~\ref{fig2}(a)-(c) by plotting three chosen banstructures corresponding to $\alpha=0.35, 0.50$ and $0.65$ respectively, with the Zak phases being mentioned along the bands. Unequivocally, we observe a redistribution in Zak phases around the 2nd order bandgap as the PhC bandstructures evolves across the nodal point at $\alpha=0.50$, establishing a topological phase transition event. With these bulk band Zak phase calculations, we aim for theoretically assigning topological identities to the associated bandgaps. To this objective, we resort to the cumulative accumulation of Zak phases as detailed in~\cite{xiao2014surface}  and ascertain the surface impedance character of the bandgaps in Fig.~\ref{fig2}. We find that in a 1D SIS binary PhC the bandgaps exhibit two complementary characters that we label in Fig.~\ref{fig2} by blue (for bandgap with negative surface impedance that we term as electric conductor-like or ENG-like) and yellow (for bandgap with positive surface impedance that we term as magnetic conductor-like or MNG-like) binary color code. Our reasoning behind this nomenclature has been detailed in Section IV and V of Supporting Information. Keeping in mind the centerstage that the Dirac equation takes in the topological studies, at this juncture, we also mention an alternative approach to mark the topological character of bandgaps based solely on the mathematical isomorphism between the Dirac equation and Maxwell's equations~\cite{tan2014photonic,huang2019observation,shi2016topological} and calculate a few relevant parameters pertaining to bandstructures of Fig.2 (a) and (c). For brevity, we only mention the outcomes here (details provided in section-V of Supporting Information)-
An abstract parameter ${m_{abst}}$ appearing in the photonic Dirac equation, can be used to classify the bandgaps in two categories: ENG-like and MNG-like bandgaps. This ${m_{abst}}$ parameter exhibits a change in sign as we transit from ENG-like bandgap to MNG-like bandgap, heralding a topological phase transition.

\subsection{Transition from theory to experiment: Bottlenecks in Measuring Topological Characteristics}\label{sec1_2}

Although by resorting to the geometric Zak phase, we can theoretically ascertain the bulk-band topology and the bandgap topological character thereof; however, the Zak phase itself does not constitute an observable in a realizable system, leading to a prevalent experimental inability in direct determination of topological character. Therefore, invariably we have to rely on indirect methods for ascertaining topological characteristics in a given sample. These methods stem from the fact that repercussions of any change in the topological character of bandgap can be seen in the dynamic phases associated with scattering parameters (reflection phase, in particular). The underlying connection here is that the surface impedance character of the bandgaps relates to the cumulative accumulation of Zak phases, which ultimately decides the reflection phase. However, experimental determination of the reflection phase is a demanding undertaking, requiring interferometric setups and coherent light sources. The situation in the case of PhCs is further aggravated by the fact that the reflection phase does not remain constant inside the bandgap; instead, it exhibits a highly dispersive behaviour inescapably ranging from $-\pi$ to $0$ or $0$ to $\pi$; therefore, determining reflection phase at one spectral point may not suffice. In most scenarios, however, we prefer to avoid these complications and employ the bulk-boundary correspondence principle to ascertain any potential difference in topological characters of two insulating samples solely based on intensity measurements (by spotting a topological surface state). In this regard, it becomes quintessential to note that such methods only permit us to ascertain the equivalence/inequivalence of bandgaps but cannot establish their absolute topological identity.

\subsection{Direct Experimental Determination of Bandgaps's Topological Character: Dispersion Measurement and Qualitative Signatures}\label{sec1_3}

 To break free from all the previously mentioned limitations, in this section, we will cultivate an experimentally amenable non-interferometric method for determining the absolute topological identity of bandgaps. Specifically, we impose the Kramers-Kronig causality considerations on the scattering response, which establish a formal connection and interdependence between amplitude and phase signatures. This permits us to measure feeble yet deterministic implications of phase dispersion and discontinuities on the intensity measurements, providing a straightforward recipe for capturing the bulk-band topology in realistic samples.

\begin{figure}[h]%
\centering
\includegraphics[width=0.5\textwidth]{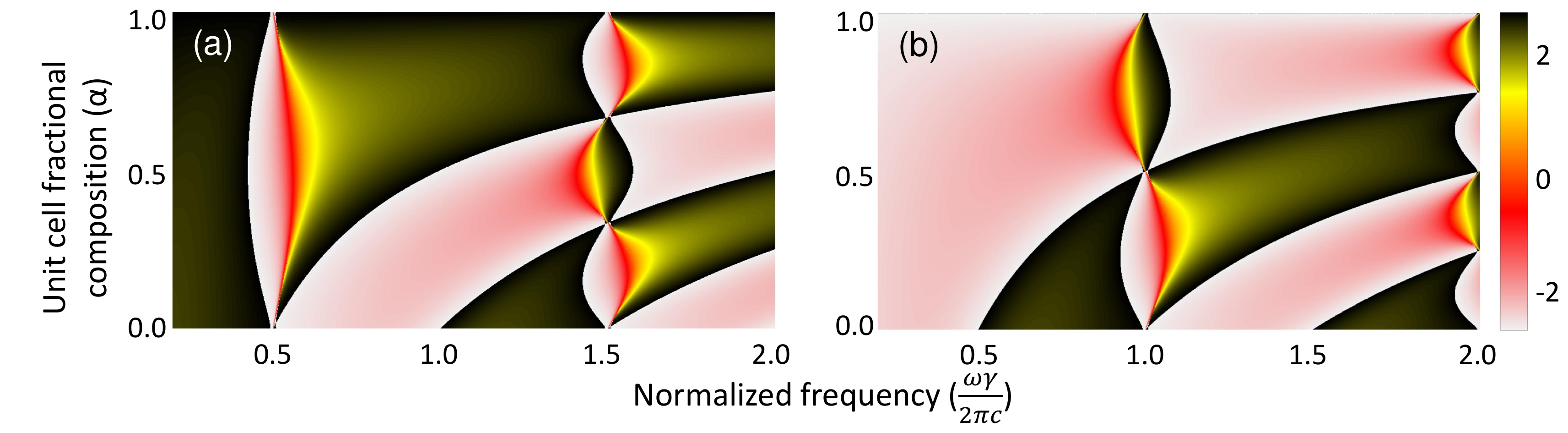}
\caption{Theoretical determination of bandgap topological identities for the semi-infinite PhC in terms of the reflection phase $\phi_n$ (rad), for (a) odd-order bandgaps; (b) even-order bandgaps, in the ($\alpha-\omega_N$) 2D parameter space.}
\label{fig3}
\end{figure}

\noindent To systemically explain the thought process, we gradually make a transition from the bulk PhC structure to a few periods realistic PhC sample. We start by relieving the periodic boundary conditions for the positive half of coordinate axis ($x>0$), leading to an air-PhC interface at $x=0$. For this semi-infinite PhC, by matching the electric and magnetic field boundary conditions at the $x=0$ interface, we calculate the dynamic phase associated with the complex reflection coefficient ${\bf r}(\omega)$ (denoted as $\phi_n$) and plot it in Fig.~\ref{fig3}(a) $\&$ (b) in the $\alpha-\omega_N$ parameter space for both even and odd order bandgaps (see Section VI of the Supporting Information for details). Focusing on the 2nd order bandgap, we see a reversal in the sign of $\phi_n$ across the Dirac point at $\alpha=0.50$ on account of the topological phase transition across $\alpha=0.50$. Apart from noticing the flipping in sign, we also note the highly dispersive nature of the $\phi_n$. Here we invoke the causal connection between the amplitude and phase response of the reflection coefficient that heralds a corresponding bearing onto the reflectance profile in case of a realistic sample. Particularly, for the linear response function $\bf{r}(\omega)$, devoid of any zeros in the upper-half of the complex $\omega$-plane, a direct interdependence exists between phase response $Arg(\bf{r}(\omega))$ and the amplitude response $\mid \bf{r}(\omega)\mid$, as stated by the Kramers-Kronig relations~\cite{gupta2022singular}:

\begin{equation}
Arg({\bf r}(\omega))=-\frac{2\omega}{\pi} PV \int_{0}^{\infty} \frac{ln \mid {\bf r}(\omega^{'})\mid}{{\omega^{'}}^2-\omega^2} d\omega^{'}
 \label{eq5}
\end{equation}

\noindent where $PV$ denotes Cauchy principal value. This relationship suggests that the distinctive phase dispersions observed in Fig.~\ref{fig3} cannot escape a manifestation in the corresponding reflectance plots.

\noindent Another headway in detecting an imprint of $\phi_n$ on the intensity measurements can be made by revisiting Fig.~\ref{fig3}(a) $\&$ (b). We notice the emergence of wrapping cut phase discontinuities (WCDs) in these phase plots across which $\phi_n$ exhibits a discontinuity of $2\pi$. These phase WCDs traverse across the parameter space and pierce through the bandgaps only at the nodal points in an oblique/asymmetric manner (we will be focusing only on 2nd order bandgaps). Their emergence must also affect the reflectance profile in an asymmetric manner. 

\begin{figure}[h]%
\centering
\includegraphics[width=0.5\textwidth]{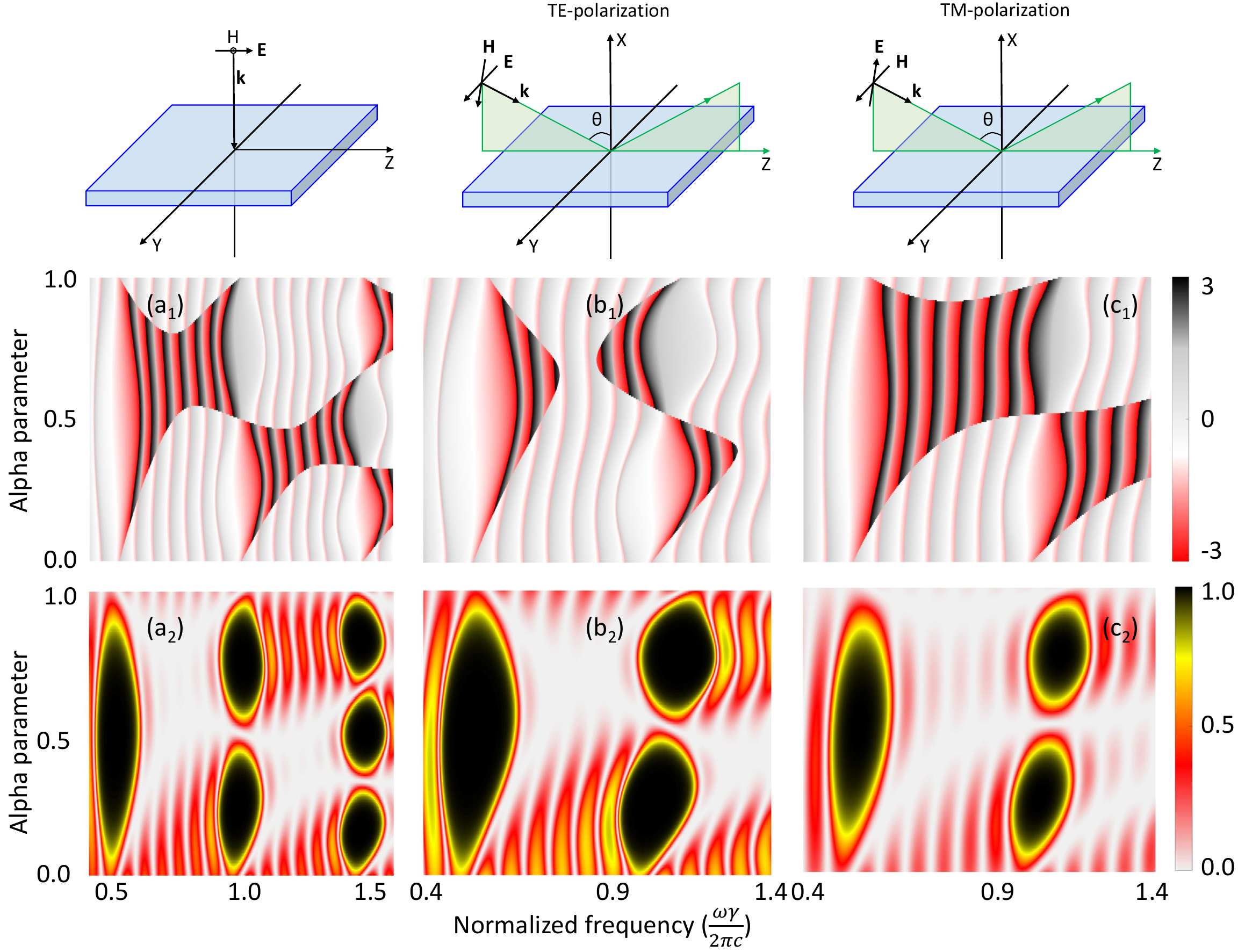}
\caption{Calculations pertaining to a finite-sized PhC: Parametric variations of reflection phase (rad) in $\alpha-\omega_N$ 2D parameter space at ($a_{1}$) normal incidence, ($b_{1}$) $45^{\circ}$ (TE-polarization), ($c_{1}$) $45^{\circ}$ (TM-polarization). Corresponding parametric variations of reflectance in $\alpha-\omega_N$ 2D parameter space at ($a_{2}$) normal incidence, ($b_{2}$) $45^{\circ}$ (TE-polarization), ($c_{2}$) $45^{\circ}$ (TM-polarization). }
\label{fig4}
\end{figure}

\noindent To verify all these propositions, we perform standard transfer matrix method (TMM) simulations for a finite size PhC sample (consisting of seven periods), and the resulting parametric variation of $\phi_n$ is plotted in Fig.~\ref{fig4}($a_{1}$). We observe that on account of differing dispersions of $\phi_n$, the two reflectance lobes in Fig.~\ref{fig4}($a_{2}$) turn out to be asymmetric. Furthermore, in consonance with our calculations of Fig.~\ref{fig3}($b$), we observe the emergence of WCDs in Fig.~\ref{fig4}($a_{1}$) as well, which pierce across the bandgaps in an asymmetric manner. These WCDs manifest themselves as oblique amplitude null lines (in fact, as a near-zero amplitude band separating the two bandgaps) in the reflectance plot of Fig.~\ref{fig4}($a_{2}$) leading to an asymmetric closing of the two bandgaps. In consequence, the prevailing symmetry in the two topologically inequivalent bandgap lobes (across the nodal point $\alpha=0.50$, as noticed in Fig.~\ref{fig1}(c)) gets broken. In other words, in an imperfectly periodic (finite-size) sample, owing to the distinctive differences in their phase response, a weak asymmetry creeps in among the two topologically inequivalent bandgap regions even in their reflectance response. However, we also notice that the asymmetry at normal incidence is very feeble and may not provide a robust experimental signature; demanding a methodology for its enhancement. For this purpose, we make use of the angularly dispersive nature of the PhC scattering response as it provides us an experimentally amenable way to alter the surface impedance (and therefore $\phi_n$). More importantly, oblique angle incidence response brings in polarization discrimination that would provide us with two measurements for each sample (in terms of two orthogonal polarization modes), materializing the possibility of absolute topological character determination. 

\noindent To witness these notions at work, we performed the calculations for $\phi_n$ at an incidence angle of $\theta=45^{\circ}$, and the results are depicted in Fig.~\ref{fig4}($b_{1}$) $\&$ ($c_{1}$) for TE and TM-polarization modes, respectively. Unequivocally, we observe polarization discrimination in these plots where the WCDs for the TE-polarization have become more oblique while those for TM-polarization have become flatter. The corresponding changes in the reflectance zero lines can be seen in Fig.~\ref{fig4}($b_{2}$) $\&$ ($c_{2}$), with a marked enhancement in the asymmetry of two reflectance lobes (corresponding to two topologically inequivalent bandgaps) for the TE-polarization . This suggests that polarization-resolved measurements may enable us to capture the distinctive topological character of two bandgaps. However, a critical caveat arises here from the fabrication and measurement viewpoint: The parametrization in terms of $\alpha$ (like the one depicted in Fig.~\ref{fig4}($b_{2}$) $\&$ ($c_{2}$)) will be an inefficient affair and will lead to serious concerns regarding the practicality of the approach. To resolve this quandary, we rely on incidence angle-resolved ($\theta$) measurements of reflectance at two fixed values of $\alpha$ ($\alpha=0.35$ and $0.65$) constituting our two investigated PhC samples, named PhC-I and PhC-II, respectively hereafter. On account of their topologically different 2nd order bandgaps and the above-explained inequivalency in reflectance, we expect their angular dispersion also to bear the very signature. Besides, since the TE and TM- polarization responses for the two bandgaps have been noticed to be different, it provides respite from the relative signature-based limited deductions, and enables us to categorize bandgaps in terms of their own polarization discriminated behaviour (bandgap absolute character determination). 

\begin{figure}[h]%
\centering
\includegraphics[width=0.5\textwidth]{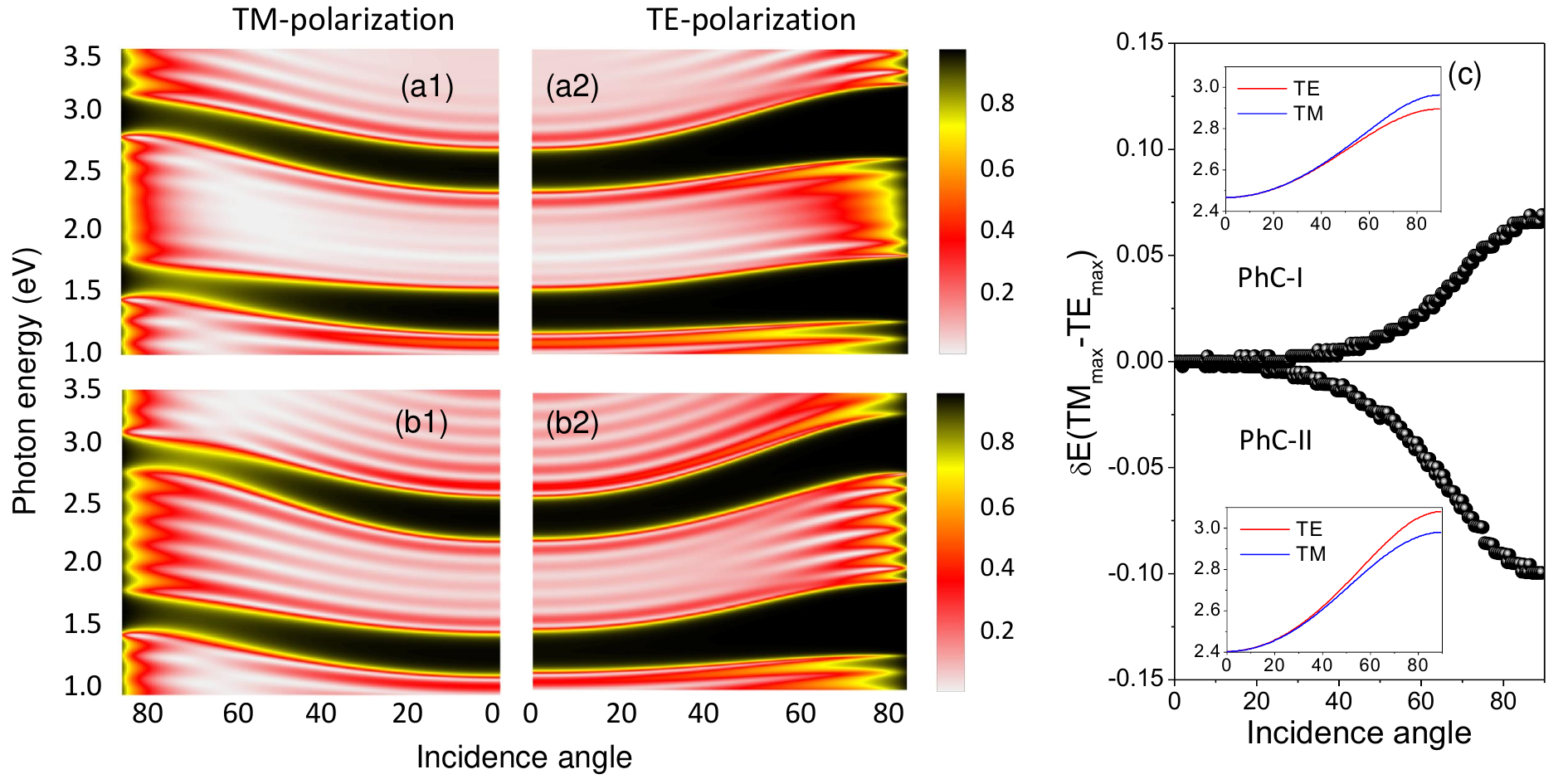}
\caption{TMM simulated polarization-resolved angular dispersion of the reflection intensity for the first and the second-order bandgaps of PhC-I: ($a_{1}$ $\&$ $a_{2}$), and PhC-II: ($b_{1}$ $\&$ $b_{2}$). In (c), we depict the extracted bandgap center dispersion (insets) and a difference parameter $\delta E(TM_{max}-TE_{max})$ for the PhC-I $\&$ II second-order bandgaps. The difference parameter $\delta E(TM_{max}-TE_{max})$ exhibits positive values for PhC-I and negative values for PhC-II.}
\label{fig5}
\end{figure}

\noindent With this premise, we again resort to the TMM simulations and obtain the angular dispersion of reflectance for the two PhC samples. The results are plotted in Fig.~\ref{fig5}($a_{1}$)-($b_{2}$). To make our observations more concrete and to prominently bring out the distinctions between the two PhC sample performances, we extract the angular dispersion of reflectance maxima (equivalently the bandgap center) from these plots and depict the results in Fig.~\ref{fig5}(c) insets. The suitability of the bandgap center for our analysis has been explained in Supporting Information Section VII. Crucially, in line with our expectations, these polarization-resolved angular dispersion measurements clearly bring out the distinctions in bandgap characters. Specifically, we see that the ENG-like bandgap of PhC-I turns out to be more responsive to TM-polarized light while the MNG-like bandgap of PhC-II turns out to be more responsive to TE-polarized light. We notice that although the signature persists throughout the span of $\theta$, the discrimination becomes more and more prominent as we approach the higher incidence angles. For the want of explicit depiction of this signature, we define a quantitative measure $\delta E(TM_{max}-TE_{max})$ and plot it in Fig.~\ref{fig5}(c) and observe that this parameter takes positive values for PhC-I and negative values for PhC-II. This categorical polarization discrimination in the dispersion constitutes our desired bandgap topological character identification signature. 

\begin{figure*}%
\centering
\includegraphics[width=1\textwidth]{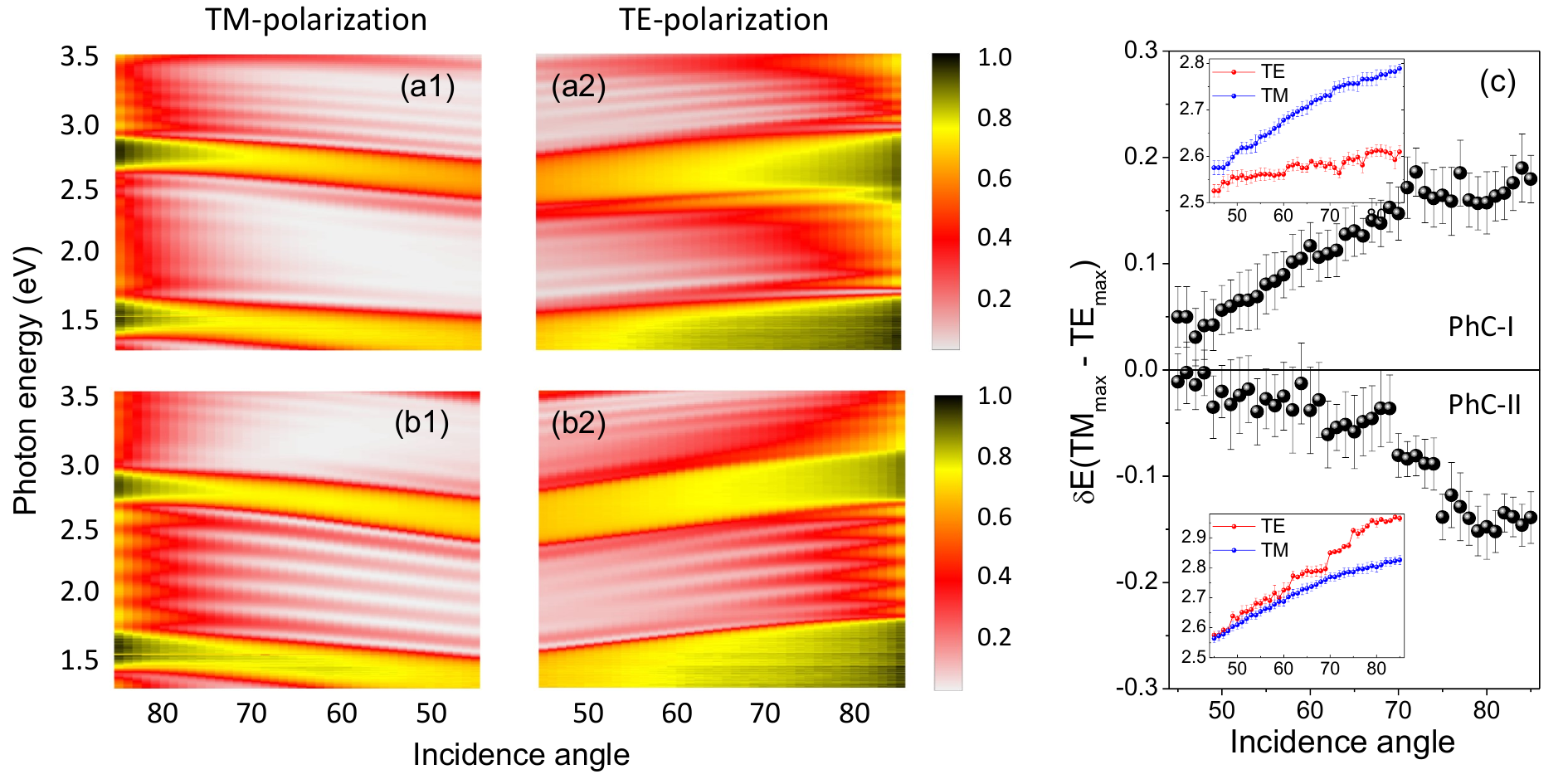}
\caption{Measured polarization-resolved angular dispersion of the reflection intensity (corresponding to one of the specimen set) for the first and the second-order bandgaps of PhC-I: ($a_{1}$ $\&$ $a_{2}$), and PhC-II: ($b_{1}$ $\&$ $b_{2}$). In (c), we depict the measured bandgap center dispersion (insets) and a difference parameter $\delta E(TM_{max}-TE_{max})$ for the PhC-I $\&$ II second-order bandgaps (these line plots depict the mean values for all three specimen sets). The difference parameter $\delta E(TM_{max}-TE_{max})$ exhibits positive values for PhC-I and negative values for PhC-II.}
\label{fig6}
\end{figure*}

\noindent For experimental validation, we have fabricated three sets of specimens (each set further consisting of two samples: PhC-I and PhC-II) on BK7 glass substrates using an ion assisted electron-beam evaporation system. After that, the polarization-resolved angular dispersion of reflectance has been measured for all the PhC samples in the angular range of $45^{\circ}-85^{\circ}$ (i.e., at higher incidence angles); for brevity, we depict the dispersion plots for only one of the sets in Fig.~\ref{fig6}($a_{1}$)-($b_{2}$). In order to adequately manifest the distinctions in the PhC topological characters, we again resort to the bandgap center angular dispersion and its mean values over all three specimen sets have been plotted in Fig.~\ref{fig6}(c) for the second-order bandgap. A direct correspondence between these experimental results and the simulated results of Fig.~\ref{fig5} can be seen that validates our propositions. Besides this, to visibly depict the ramifications of this difference in angular dispersion of two topologically distinct PhC samples onto their spectral response, in Fig.~4 of Supporting Information, we also plot the measured spectral performance for PhC-I and PhC-II, at a large enough incidence angle of $70^{\circ}$. 

\begin{figure}[h]%
\centering
\includegraphics[width=0.5\textwidth]{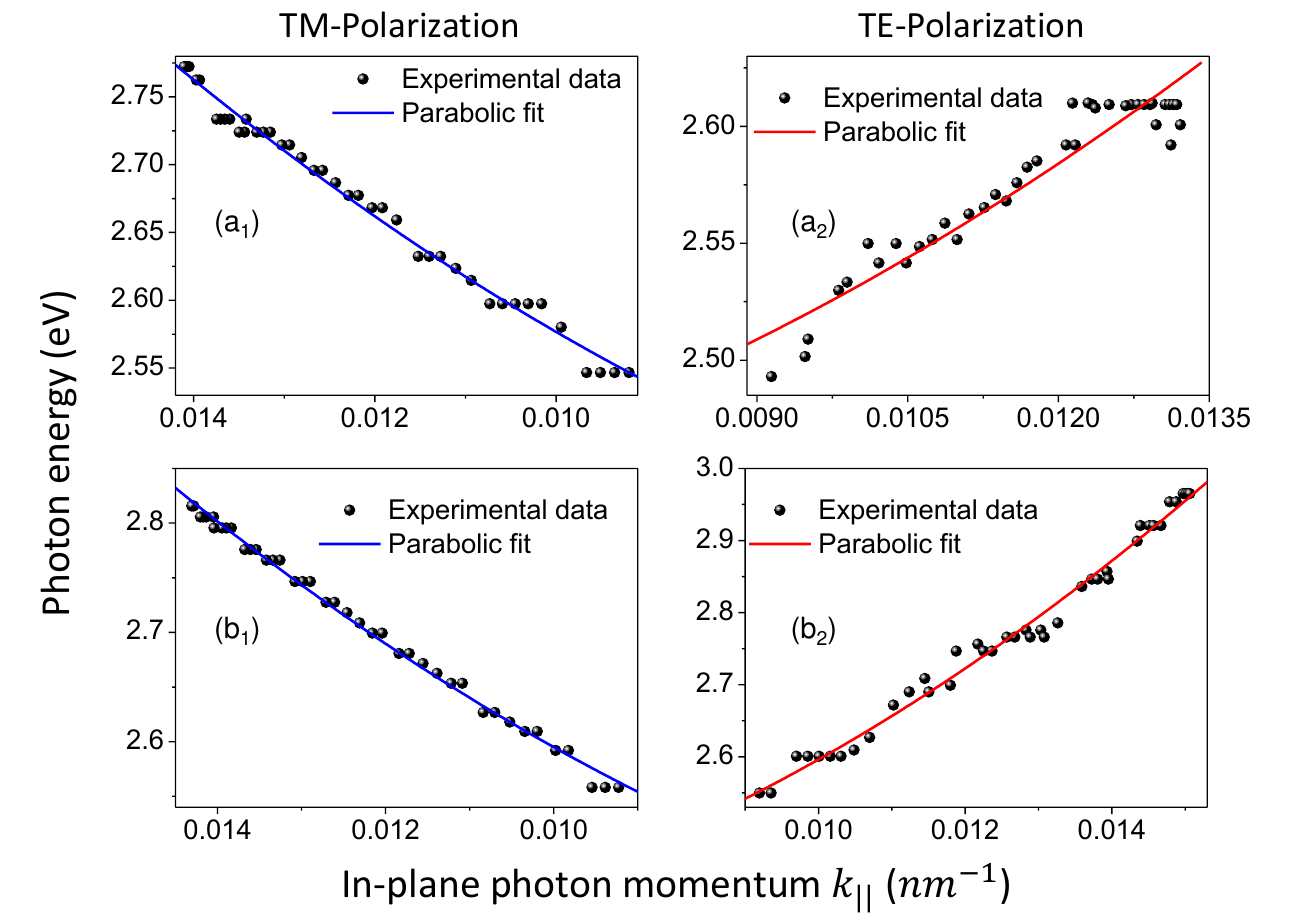}
\caption{Measured polarization-resolved bandgap center dispersions ($E-k_{\mid \mid}$ or energy-momentum diagrams) for the second-order bandgap, fitted with the proposed analytical model, for PhC-I ($a_{1}$,$a_{2}$) $\&$ PhC-II ($b_{1}$,$b_{2}$). }
\label{fig7}
\end{figure}

\subsection{Quantitative Criterion: Differential Effective Mass and an Experimentally Discernible Classifier for Topological Order Characterization}\label{sec1_4}
In the previous section, we discovered the qualitative signatures for determining the bandgap's absolute topological identity. In order to further concretize our findings, here we put forward a quantitative criterion that furnishes us with a direct and straightforward experimental marker of bandgap topological character. 

\noindent From TMM simulations of Fig.~\ref{fig5}, we observe that the bandgap center exhibits a squared sinusoidal-like dispersion with incidence angle $\theta$. In order to extract certain pragmatic quantities from this dispersion, we will recast it in the standard energy-momentum formalism. We begin by invoking translational invariance of investigated samples that leads to the conservation of the in-plane component of the photon momentum ($k_{\mid \mid}=k_z$). In the planar geometry of our samples, $k_{\mid \mid}$ has a direct dependence on $\theta$ ($k_{\mid \mid}=k_0sin\theta$); therefore, we recognize that angular dispersion measurements are capable of extracting the information pertaining to the projected band diagram, facilitating an experimentally conducive method for capturing the bandgap behaviour. To explicitly highlight the nature of bandgap-center dispersion, in Fig.~5 of Supporting Information-Section IX, we plot the recasted $E-k_{\mid \mid}$ dispersion diagrams for the two PhCs. 

\noindent In the next section, we will derive the dispersion expression for the investigated scenarios, which takes the following form:
\begin{equation}
    E=\sqrt{E_{0}^{2} + E_{1}^{2}k^{2}_{\mid \mid}}
     \label{eq6}
\end{equation}
\noindent where $k_{\mid \mid}=k_{0}sin\theta$, $E_{0}$ is the photon energy corresponding to bandgap center at incidence angle $\theta=0$, and $E_{1}$ characterizes the rate of change of dispersion curve with respect to $k_{\mid \mid}$. While the parameter $E_0$ remains same for both TE $\&$ TM polarizations, $E_1$ takes on different values for TE and TM-mode dispersions, and hence can serve as a quantitative measure of polarization discriminated response. Under the condition $E_{0} > E_{1}$ (which prevails in our samples), the above dispersion relation can be simplified to a parabolic form:

\begin{equation}
    E \approx E_{0} + \frac{E^{2}_{1}}{2E_{0}}k^{2}_{\mid \mid}
     \label{eq7}
\end{equation} 

\noindent We fit this parabolic dispersion relation to the previously mentioned dispersions of Supporting Information Fig.~5 and the results are plotted in Fig.~6 of Supporting Information, depicting an excellent match between the two curves. For outright confirmation, we fit the parabolic dispersion relation of Eq.~\ref{eq7}, to the experimentally obtained data as well (for one of the specimen set) and the results are plotted in Fig.~\ref{fig7}, demonstrating a satisfactory match with maximum root mean square error being $1.4\%$. The obtained fitting parameters are provided in Table-II of Supporting Information, confirming the parabolic nature of all the measured dispersions. This observation is important as it assigns a constant curvature to the dispersion curve, allowing us to invoke a parameter in congruence with the standard definition of effective mass for non-relativistic particles (inverse of $E-p$ diagram curvature, $p$ being the momentum). Specifically, we define an in-plane effective mass parameter (for the bandgap center dispersion) in terms of the curvature of the fitted dispersion relation of Eq.~\ref{eq7} :
\begin{equation}
    m_{eff} = \left(\frac{d^{2}E}{dp^{2}_{\mid \mid}}\right)^{-1} = \frac{E_0}{E^2_1}
     \label{eq8}
\end{equation}

\noindent Evidently, we observe that, $m_{eff}$ turns out to be a constant for a given sample and for a particular polarization while being proportional to the energy of the bandgap center ($E_{0}$) at $\theta=0$ and inversely proportional to the $E^{2}_{1}$. The values of $m_{eff}$ has been calculated for the two PhCs which turns out to be about $1.9376\times10^{-5}m_{e}$ for the PhC-I TE-mode and about $1.7648\times10^{-5}m_{e}$ for the PhC-I TM-mode. Similarly, the $m_{eff}$ values for PhC-II are $1.3908\times10^{-5}m_{e}$ for the TE-mode and $1.5233\times10^{-5}m_{e}$ for the TM-mode, where $m_{e}$ denotes the mass of an electron. With these numbers, we are now in a position to enumerate a quantitative criterion for direct ascertainment of bandgap topological character: 
(1) Bandgap is ENG-like if $(m_{eff})_{TM}<(m_{eff})_{TE}$
(2) Bandgap is MNG-like if $(m_{eff})_{TM}>(m_{eff})_{TE}$.

\noindent It is important to mention here that both the values $(m_{eff})_{TM}$ $\&$ $(m_{eff})_{TE}$ correspond to a single PhC, ergo, the above-mentioned criteria is capable of uncovering the absolute topological identity of bandgaps. Moving further, we work towards bringing our criteria directly in commensuration to the Dirac-equation formalism, where the sign of effective mass foretells the distinction between particle and antiparticle wavefunctions. To this aim, we define an in-plane differential effective mass parameter ($\delta m$) in the following manner:
\begin{equation}
    \delta m=(m_{eff})_{TM} - (m_{eff})_{TE}
     \label{eq9}
\end{equation}

\noindent This parameter can exhibit both negative and positive values depending upon whether $(m_{eff})_{TM}<(m_{eff})_{TE}$ or vice-versa.  In the present context, we are dominantly concerned about the sign information of $\delta m$ with: $\delta m<0$ denoting an ENG-like bandgap, and $\delta m>0$ denoting a MNG-like bandgap. At this juncture, we also mention that relying on these distinctions of bandgap response in terms of sign information of measurable $\delta m$ parameter, we can invoke the relationship between the evolution of accumulated Zak phase and the surface property of the bandgap. This facilitates an operational method where we can ascertain the Zak phase of any $n^{\text{th}}$ bulk-band by performing the polarization-resolved bandgap center dispersion measurements for the $n^{\text{th}}$ and $(n-1)^{\text{th}}$ bandgaps (and finding the $sgn(\delta m^{(n)})~ \& ~sgn(\delta m^{(n-1)})$ thereafter). To exemplify this approach of ascertainment of Zak phase, in Section X of Supporting Information, we find out the Zak phases for the second bands of both, $\alpha=0.35~ \& \alpha=0.65$ PhC samples and also provide a general expression for obtaining the Zak phase of any $n^{\text{th}}$ bulk-band.

\noindent We also want to mention that for the objective of bandgap topological character determination although we have kept our focus on the sign information of $\delta m$ but the magnitude of $\delta m$ becomes equally important when we intend to employ these ENG-like/MNG-like bandgaps in some application, such as engendering a topological surface state. Specifically, if we intend to engineer the dispersion and other properties pertaining to the engendered topological surface state, the parameter $\delta m=(m_{eff})_{TM} - (m_{eff})_{TE}$ becomes critically important. Therefore, by quantitatively characterizing the polarization discriminated dispersion response of PhC samples in terms of $\delta m$ parameter, our manuscript initiates an important step where it provides not only a simple framework for bandgap character determination (in terms of the sign of  $\delta m$ that denotes the nature of polarization splitting) but also enables us to obtain/characterize engineered dispersions for the ENG-like and MNG-like bandgaps and ensuing surafce states (in terms of the magnitude of $\delta m$ that denotes the amount of polarization splitting).

\noindent Finally, we encapsulate this topological character information in a quantized parameter $\chi$, thereby, conceiving an experimentally discernible bandgap topological character classifier for SIS binary 1D PhC, in the following manner:

\begin{equation}
    \chi=sgn[\delta m]= \frac{\delta m}{\mid \delta m \mid}
     \label{eq10}
\end{equation}

\noindent Evidently, the classifier $\chi$ can take two values: $\chi=+1$ for a MNG-like bandgap, and $\chi=-1$ for an ENG-like bandgap. Notice that, all these parameters have been worked out for both the simulated and experimental dispersion curves of Fig.~\ref{fig5} $\&$ \ref{fig6} and congruent results have been obtained which are compiled in Table-I $\&$ Table-II of Supporting Information.  Also, in order to witness the explicit physical implications of classifier $\chi$, we can probe the PhC samples in the attenuated total reflection configuration, where $\chi=+1$ and $\chi=-1$ samples show complementarity in terms of polarization-resolved existence of the high-k propagating surface states. Specifically, the propagating surface state will appear for TE-polarization in case of $\chi=+1$ PhC, on the other hand, it will appear for TM-polarization for $\chi=-1$ PhC.  
 
\noindent We conclude this section by mentioning that Eq.~\ref{eq9}, $\&$ \ref{eq10} summarize one of the main outcomes of our work, where we present a new direct experimental identifier of absolute topological identities of photonic bandgaps in terms of $\chi$ or the sign of $\delta m$. 

\subsection{Capturing the Topological Signatures in PhC Bandgap Angular Dispersion with a Semi-Analytical Model}\label{sec1_5}
This section pertains to the development of a simple mathematical model that can assist our understanding of the obtained results. To this objective, we work with a PhC effective medium (effective parameters calculations provided in Section V of Supporting Information) and derive the polarization-resolved dispersion relations.  
Although we are investigating a 1D system where at normal incidence, a two-component field vector suffices but to arrive at the simulated configuration of Fig.~\ref{fig5}, we must include the possibility of two orthogonal polarization modes in our analysis. To that objective, we write the electromagnetic (EM) mode as a four-components E and H field vector of the form $\Psi=\left(E_{x}, E_{z}, H_{x}, H_{z} \right)^{T}$ and also write the constitutive parameters for the effective PhC medium in a tensorial form to arrive at the following wave equation (For detailed derivation refer to Section XI Supporting Information):

\begin{widetext} 
\begin{equation}
     \begin{pmatrix}
      -\frac{\partial^{2}}{\partial z^{2}} & \frac{\partial^{2}}{\partial z \partial x} & 0 & 0 \\
      
      \frac{\partial^{2}}{\partial z \partial x}  & -\frac{\partial^{2}}{\partial x^{2}} & 0 & 0 \\
     0 & 0 &  -\frac{\partial^{2}}{\partial z^{2}} & \frac{\partial^{2}}{\partial z \partial x} \\
     0 & 0 & \frac{\partial^{2}}{\partial z \partial x}  & -\frac{\partial^{2}}{\partial x^{2}} \\
    \end{pmatrix} \Psi = \omega^{2}\epsilon_{0}\mu_{0}\begin{pmatrix}
         \epsilon_{x} & \epsilon_{xz}  & 0 & 0 \\
         \epsilon^{*}_{xz} & \epsilon_{z}  & 0 & 0 \\
         0 & 0 & \mu_{x} & \mu_{xz} \\
         0 & 0 & \mu^{*}_{xz} & \mu_{z} \\
     \end{pmatrix} \Psi
      \label{eq11}
\end{equation}
\end{widetext} 

\noindent which promulgates the below-mentioned dispersion relations for the TE-mode and TM-modes respectively:

\begin{equation}
    \frac{\omega^{2}}{c^{2}}\mu_{x}\mu_{z} - k^{2}_{x}\mu_{x}- k^{2}_{z}\mu_{z} =  \frac{\omega^{2}}{c^{2}} \mid \mu_{xz} \mid^{2} + k_{x}k_{z}\left( \mu_{xz} + \mu^{*}_{xz} \right)
     \label{eq12}
\end{equation}

\begin{equation}
    \frac{\omega^{2}}{c^{2}}\epsilon_{x}\epsilon_{z} - k^{2}_{x}\epsilon_{x}- k^{2}_{z}\epsilon_{z} =  \frac{\omega^{2}}{c^{2}} \mid \epsilon_{xz} \mid^{2} + k_{x}k_{z}\left( \epsilon_{xz} + \epsilon^{*}_{xz} \right)
     \label{eq13}
\end{equation}

\noindent  Next, we conceive a vacuum-PhC medium interface and couple an incident plane wave with its in-plane photon momentum being conserved. Under these conditions the above dispersion relations convert to:
\begin{equation}
     \frac{\omega^{2}}{c^{2}} =   \frac{\epsilon_{eff}\mu_{eff}}{\mu_{z}}k^{2}_{0} +   \left( \frac{1}{\mu_{x}} - \frac{1}{\mu_{z}} \right) k^{2}_{0}sin^{2}(\theta_{0})
      \label{eq14}
\end{equation} for the TE-mode and 
\begin{equation}
     \frac{\omega^{2}}{c^{2}} =   \frac{\epsilon_{eff}\mu_{eff}}{\epsilon_{z}}k^{2}_{0} +   \left( \frac{1}{\epsilon_{x}} - \frac{1}{\epsilon_{z}} \right) k^{2}_{0}sin^{2}(\theta_{0}) 
      \label{eq15}
\end{equation} for the TM-mode (notice here that for simplicity we have assumed the cross-coupling terms $\epsilon_{xz}$ and $\mu_{xz}$ to be zero).

\noindent By employing appropriate values of constitutive parameters, we have plotted these dispersion relations in Fig.~8 of Supporting Information for both the PhC-I $\&$ PhC-II. From these plots, we obtain a qualitatively similar behaviour for bandgap dispersion, as presented in Fig.~\ref{fig5}.

\subsection{Robustness of the Conceived Signatures to the Variations in System Size}\label{sec3}

\begin{figure*}%
\centering
\includegraphics[width=0.8\textwidth]{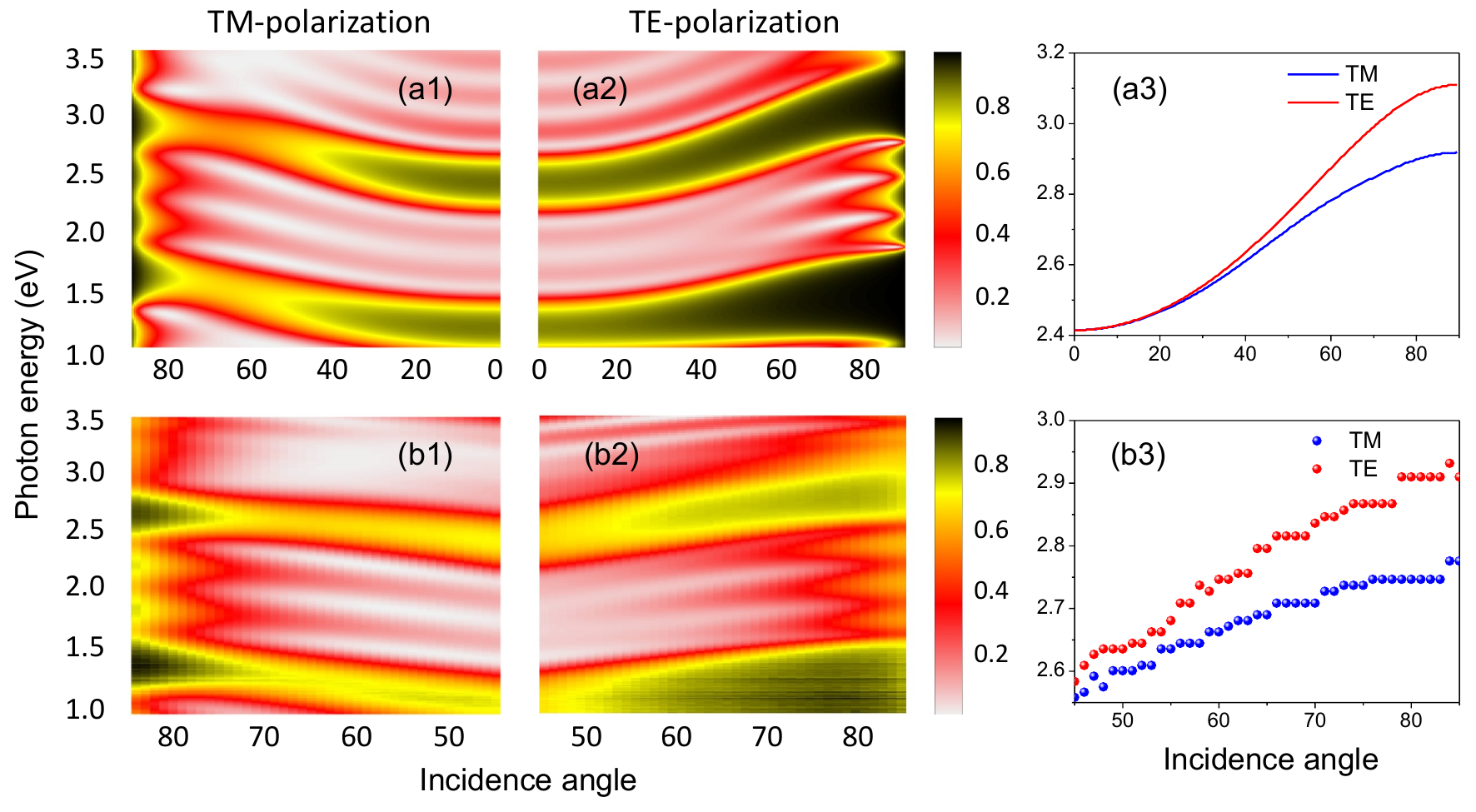}
\caption{($a_{1}$) $\&$ ($a_{2}$): TMM simulated polarization-resolved angular dispersions of the reflection intensity for the first and the second-order bandgaps of a reduced-sized PhC (for $\alpha=0.65$ and consisting of four unit cells).  ($b_{1}$) $\&$ ($b_{2}$): Corresponding measurement results for the reduced-sized PhC sample. ($a_{3}$) and ($b_{3}$) represent the simulated and measured bandgap center dispersions for the second-order bandgap, respectively.}
\label{fig9}
\end{figure*} 

\noindent While studying topologically non-trivial systems, a question arises: do we need to incorporate a large number of periods to see a proper correspondence with the bulk behavior? It turns out that on account of the quantized nature of topological invariants, the topological properties of bulk bandstructure can manifest themselves in systems with arbitrarily small sizes. In previous sections, we have demonstrated that sufficient correspondence prevails between bulk behaviour and PhC samples consisting only of seven periods. Here, we further establish the generality of all our propositions on a sample where we further reduce the number of periods down to only four periods (for $\alpha=0.65$ case) and obtain its polarization-resolved angular dispersion. Both the calculation and measurement results for this new sample are provided in Fig.~\ref{fig9}. We observe that not only the signature of the bandgap topological character for this MNG-like bandgap remains the same, but we also find near-perfect overlapping of results to those in Fig.~\ref{fig5}(c). This remarkable robustness of bandcenter dispersion to the variation in system size further points toward the fundamental nature of the investigated signature. The measurement results for this sample are plotted in Fig.~\ref{fig9}($b_{1}$)-($b_{3}$), which, again, demonstrate excellent correspondence with simulations. 

\noindent As a final verification of our propositions, we have also performed the calculations for $\delta E(TM_{max}-TE_{max})$ across the span of $\alpha$ and the resultant two-dimensional plot (in the $\alpha-\theta$ parameter space) is presented in Section XIII of the Supporting Information file. From this plot, we can confirm that the proposed signature (in terms of $\delta E(TM_{max}-TE_{max})$) is not $\alpha$ specific but works for all values of $\alpha$ (except near the points where the bandgap itself closes). Furthermore, we also mention here that our framework exposes a fundamental distinction between the two complementary bandgaps, hence remains equally valid for even-ordered as well as odd-ordered bandgpas.

\section{Conclusion}\label{sec13}

To summarize, our work proposes $\&$ demonstrates (at optical frequencies) an experimentally conducive and robust method of determining the bandgaps' absolute topological character. The genesis of our approach lies in Kramers- Kronig causality considerations which forge an interdependence between the amplitude and phase responses, permitting us to collect sufficient information regarding the changes occurring in Bloch eigenfunction's spatial distributions directly from parametric reflectance plots. Specifically, the differing phase dispersions of topologically distinct bandgaps supplemented with the presence of wrapping cut discontinuities in the parametric phase plots leave their imprints on the reflectance measurements that we have captured using the polarization-resolved angular dispersion measurements. On the basis of these dispersions, we theoretically and experimentally establish that the bandgaps of 1D PhC exhibit two complementary characters: one being more responsive to the TM-polarized light (ENG-like character) and the other being more responsive to the TE-polarized light (MNG-like character). Harboring on these discernible signatures, we define a differential effective mass parameter and a bandgap classifier thereof, which encapsulate the bandgap dispersion information and serve as appropriate markers of topological identity. Furthermore, it also engenders the possibilities of dispersion engineering of the topological surface states. We hope our results provide a new perspective regarding topologically non-trivial systems and help circumvent some of the difficulties afflicting the experimental investigations of photonic topological insulators.


\section*{Methods}
\subsubsection*{Photonic Crystal Fabrication}
The PhC samples were prepared on BK7 glass substrates using the ion assisted electron-beam evaporation system (Evatec AG: BAK761). Before fabrication, the substrates were cleaned as per the standard procedure. Thereafter, the substrates were loaded into the vacuum chamber, and a base pressure of $5\times10^{-6}$ mbar was attained.  The deposition of $SiO_2$ and $TiO_2$ was performed at a working pressure of $2\times10^{-4}$ mbar with argon and oxygen flow rates of $20$~sccm and $8$~sccm, respectively. The ion source current was kept at $3.2$ Amps. at a potential difference of $120$ Volts, and the substrate temperature was maintained at $150^{\circ}C$. Under these conditions, the deposition rates of $0.5$~nm/sec and $0.2$~nm/sec have been obtained for $SiO_2$ and $TiO_2$ (measured from quartz crystal thickness monitor). With these fabrication parameters, we have prepared three specimen sets (each set further consisting of two PhC samples corresponding to PhC-I \& PhC-II designs), comprising seven unit cells each. Another PhC sample consisting of four unit cells has been prepared under identical experimental conditions to demonstrate the robustness of proposed signatures. 

\subsubsection*{Polarization-Resolved Angular Dispersion Measurements}
The polarization-resolved reflectance for the PhC samples was measured using the spectroscopic ellipsometer (J. A. Woollam, M-2000) in the angular range of $45^{\circ}$ to $85^{\circ}$ deg. The broadband light source of the ellipsometer allows us to measure the relative reflection intensity from the PhC samples in a spectral range of $350$~nm to $1000$~nm. For all the measurements, we have employed an angular resolution of $1^{\circ}$ deg and a spectral resolution of $1.57$~nm.

\subsubsection*{Spectroscopic Reflectance Measurements for concatenated sample}
These measurements were performed to spot the presence of topological surface state on the PhC-I, and PhC-II concatenated sample. The measurements were performed with unpolarized light at near-normal incidence ($10^{\circ}$) using the spectrophotometer (Agilent Cary 5000 with UMA) in a broad spectral range of $350-2000$~nm. The spectral resolution of $1$~nm has been employed for these measurements.

{\bf Supporting Information:} Supporting Information.\\ 

\section*{Declarations}

\section*{Conflict of interest/Competing interests}
The authors declare no competing interests.

\section*{Authors' contributions}
N.K.G. conceived the presented idea and prepared the roadmap. N.K.G performed the simulations and finalized the designs. S.S. verified and discussed the simulation results. M.K. fabricated the samples. N.K.G. performed the polarization-resolved measurements. N.K.G. and S.S. developed the semi-analytical model. A.K.T. vindicated all the simulations and theoretical calculations. N.K.G. wrote the manuscript with significant contributions from S.S. and A.K.T. The findings of this work are supervised by S.S.P., H.W., and S.A.R. who have also provided their critical inputs for improvement. All authors discussed the results and reviewed the final manuscript.

\bibliography{references_aps}

\begin{thebibliography}{61}%
\makeatletter
\providecommand \@ifxundefined [1]{%
 \@ifx{#1\undefined}
}%
\providecommand \@ifnum [1]{%
 \ifnum #1\expandafter \@firstoftwo
 \else \expandafter \@secondoftwo
 \fi
}%
\providecommand \@ifx [1]{%
 \ifx #1\expandafter \@firstoftwo
 \else \expandafter \@secondoftwo
 \fi
}%
\providecommand \natexlab [1]{#1}%
\providecommand \enquote  [1]{``#1''}%
\providecommand \bibnamefont  [1]{#1}%
\providecommand \bibfnamefont [1]{#1}%
\providecommand \citenamefont [1]{#1}%
\providecommand \href@noop [0]{\@secondoftwo}%
\providecommand \href [0]{\begingroup \@sanitize@url \@href}%
\providecommand \@href[1]{\@@startlink{#1}\@@href}%
\providecommand \@@href[1]{\endgroup#1\@@endlink}%
\providecommand \@sanitize@url [0]{\catcode `\\12\catcode `\$12\catcode
  `\&12\catcode `\#12\catcode `\^12\catcode `\_12\catcode `\%12\relax}%
\providecommand \@@startlink[1]{}%
\providecommand \@@endlink[0]{}%
\providecommand \url  [0]{\begingroup\@sanitize@url \@url }%
\providecommand \@url [1]{\endgroup\@href {#1}{\urlprefix }}%
\providecommand \urlprefix  [0]{URL }%
\providecommand \Eprint [0]{\href }%
\providecommand \doibase [0]{https://doi.org/}%
\providecommand \selectlanguage [0]{\@gobble}%
\providecommand \bibinfo  [0]{\@secondoftwo}%
\providecommand \bibfield  [0]{\@secondoftwo}%
\providecommand \translation [1]{[#1]}%
\providecommand \BibitemOpen [0]{}%
\providecommand \bibitemStop [0]{}%
\providecommand \bibitemNoStop [0]{.\EOS\space}%
\providecommand \EOS [0]{\spacefactor3000\relax}%
\providecommand \BibitemShut  [1]{\csname bibitem#1\endcsname}%
\let\auto@bib@innerbib\@empty
\bibitem [{\citenamefont {Kim}\ \emph {et~al.}(2020)\citenamefont {Kim},
  \citenamefont {Jacob},\ and\ \citenamefont {Rho}}]{kim2020recent}%
  \BibitemOpen
  \bibfield  {author} {\bibinfo {author} {\bibfnamefont {M.}~\bibnamefont
  {Kim}}, \bibinfo {author} {\bibfnamefont {Z.}~\bibnamefont {Jacob}},\ and\
  \bibinfo {author} {\bibfnamefont {J.}~\bibnamefont {Rho}},\ }\bibfield
  {title} {\bibinfo {title} {Recent advances in 2d, 3d and higher-order
  topological photonics},\ }\href@noop {} {\bibfield  {journal} {\bibinfo
  {journal} {Light: Science \& Applications}\ }\textbf {\bibinfo {volume}
  {9}},\ \bibinfo {pages} {1} (\bibinfo {year} {2020})}\BibitemShut {NoStop}%
\bibitem [{\citenamefont {Cheng}\ \emph {et~al.}(2022)\citenamefont {Cheng},
  \citenamefont {Wang}, \citenamefont {Ke}, \citenamefont {Chen}, \citenamefont
  {Yu}, \citenamefont {Kivshar}, \citenamefont {Lee},\ and\ \citenamefont
  {Pan}}]{cheng2022asymmetric}%
  \BibitemOpen
  \bibfield  {author} {\bibinfo {author} {\bibfnamefont {Q.}~\bibnamefont
  {Cheng}}, \bibinfo {author} {\bibfnamefont {H.}~\bibnamefont {Wang}},
  \bibinfo {author} {\bibfnamefont {Y.}~\bibnamefont {Ke}}, \bibinfo {author}
  {\bibfnamefont {T.}~\bibnamefont {Chen}}, \bibinfo {author} {\bibfnamefont
  {Y.}~\bibnamefont {Yu}}, \bibinfo {author} {\bibfnamefont {Y.~S.}\
  \bibnamefont {Kivshar}}, \bibinfo {author} {\bibfnamefont {C.}~\bibnamefont
  {Lee}},\ and\ \bibinfo {author} {\bibfnamefont {Y.}~\bibnamefont {Pan}},\
  }\bibfield  {title} {\bibinfo {title} {Asymmetric topological pumping in
  nonparaxial photonics},\ }\href@noop {} {\bibfield  {journal} {\bibinfo
  {journal} {Nature Communications}\ }\textbf {\bibinfo {volume} {13}},\
  \bibinfo {pages} {1} (\bibinfo {year} {2022})}\BibitemShut {NoStop}%
\bibitem [{\citenamefont {Tang}\ \emph {et~al.}(2022)\citenamefont {Tang},
  \citenamefont {He}, \citenamefont {Shi}, \citenamefont {Liu}, \citenamefont
  {Chen},\ and\ \citenamefont {Dong}}]{tang2022topological}%
  \BibitemOpen
  \bibfield  {author} {\bibinfo {author} {\bibfnamefont {G.-J.}\ \bibnamefont
  {Tang}}, \bibinfo {author} {\bibfnamefont {X.-T.}\ \bibnamefont {He}},
  \bibinfo {author} {\bibfnamefont {F.-L.}\ \bibnamefont {Shi}}, \bibinfo
  {author} {\bibfnamefont {J.-W.}\ \bibnamefont {Liu}}, \bibinfo {author}
  {\bibfnamefont {X.-D.}\ \bibnamefont {Chen}},\ and\ \bibinfo {author}
  {\bibfnamefont {J.-W.}\ \bibnamefont {Dong}},\ }\bibfield  {title} {\bibinfo
  {title} {Topological photonic crystals: Physics, designs, and applications},\
  }\href@noop {} {\bibfield  {journal} {\bibinfo  {journal} {Laser \& Photonics
  Reviews}\ }\textbf {\bibinfo {volume} {16}},\ \bibinfo {pages} {2100300}
  (\bibinfo {year} {2022})}\BibitemShut {NoStop}%
\bibitem [{\citenamefont {Yang}\ \emph {et~al.}(2017)\citenamefont {Yang},
  \citenamefont {Guo}, \citenamefont {Tremain}, \citenamefont {Barr},
  \citenamefont {Gao}, \citenamefont {Liu}, \citenamefont {B{\'e}ri},
  \citenamefont {Xiang}, \citenamefont {Fan}, \citenamefont {Hibbins} \emph
  {et~al.}}]{yang2017direct}%
  \BibitemOpen
  \bibfield  {author} {\bibinfo {author} {\bibfnamefont {B.}~\bibnamefont
  {Yang}}, \bibinfo {author} {\bibfnamefont {Q.}~\bibnamefont {Guo}}, \bibinfo
  {author} {\bibfnamefont {B.}~\bibnamefont {Tremain}}, \bibinfo {author}
  {\bibfnamefont {L.~E.}\ \bibnamefont {Barr}}, \bibinfo {author}
  {\bibfnamefont {W.}~\bibnamefont {Gao}}, \bibinfo {author} {\bibfnamefont
  {H.}~\bibnamefont {Liu}}, \bibinfo {author} {\bibfnamefont {B.}~\bibnamefont
  {B{\'e}ri}}, \bibinfo {author} {\bibfnamefont {Y.}~\bibnamefont {Xiang}},
  \bibinfo {author} {\bibfnamefont {D.}~\bibnamefont {Fan}}, \bibinfo {author}
  {\bibfnamefont {A.~P.}\ \bibnamefont {Hibbins}}, \emph {et~al.},\ }\bibfield
  {title} {\bibinfo {title} {Direct observation of topological surface-state
  arcs in photonic metamaterials},\ }\href@noop {} {\bibfield  {journal}
  {\bibinfo  {journal} {Nature Communications}\ }\textbf {\bibinfo {volume}
  {8}},\ \bibinfo {pages} {1} (\bibinfo {year} {2017})}\BibitemShut {NoStop}%
\bibitem [{\citenamefont {Ke}\ \emph {et~al.}(2016)\citenamefont {Ke},
  \citenamefont {Qin}, \citenamefont {Mei}, \citenamefont {Zhong},
  \citenamefont {Kivshar},\ and\ \citenamefont {Lee}}]{ke2016topological}%
  \BibitemOpen
  \bibfield  {author} {\bibinfo {author} {\bibfnamefont {Y.}~\bibnamefont
  {Ke}}, \bibinfo {author} {\bibfnamefont {X.}~\bibnamefont {Qin}}, \bibinfo
  {author} {\bibfnamefont {F.}~\bibnamefont {Mei}}, \bibinfo {author}
  {\bibfnamefont {H.}~\bibnamefont {Zhong}}, \bibinfo {author} {\bibfnamefont
  {Y.~S.}\ \bibnamefont {Kivshar}},\ and\ \bibinfo {author} {\bibfnamefont
  {C.}~\bibnamefont {Lee}},\ }\bibfield  {title} {\bibinfo {title} {Topological
  phase transitions and thouless pumping of light in photonic waveguide
  arrays},\ }\href@noop {} {\bibfield  {journal} {\bibinfo  {journal} {Laser \&
  Photonics Reviews}\ }\textbf {\bibinfo {volume} {10}},\ \bibinfo {pages}
  {995} (\bibinfo {year} {2016})}\BibitemShut {NoStop}%
\bibitem [{\citenamefont {Chen}\ \emph {et~al.}(2019)\citenamefont {Chen},
  \citenamefont {Yu}, \citenamefont {Song}, \citenamefont {Yu}, \citenamefont
  {Ye}, \citenamefont {Xie}, \citenamefont {Shen}, \citenamefont {Pan},\ and\
  \citenamefont {Cheng}}]{chen2019distinguishing}%
  \BibitemOpen
  \bibfield  {author} {\bibinfo {author} {\bibfnamefont {T.}~\bibnamefont
  {Chen}}, \bibinfo {author} {\bibfnamefont {Y.}~\bibnamefont {Yu}}, \bibinfo
  {author} {\bibfnamefont {Y.}~\bibnamefont {Song}}, \bibinfo {author}
  {\bibfnamefont {D.}~\bibnamefont {Yu}}, \bibinfo {author} {\bibfnamefont
  {H.}~\bibnamefont {Ye}}, \bibinfo {author} {\bibfnamefont {J.}~\bibnamefont
  {Xie}}, \bibinfo {author} {\bibfnamefont {X.}~\bibnamefont {Shen}}, \bibinfo
  {author} {\bibfnamefont {Y.}~\bibnamefont {Pan}},\ and\ \bibinfo {author}
  {\bibfnamefont {Q.}~\bibnamefont {Cheng}},\ }\bibfield  {title} {\bibinfo
  {title} {Distinguishing the topological zero mode and tamm mode in a
  microwave waveguide array},\ }\href@noop {} {\bibfield  {journal} {\bibinfo
  {journal} {Annalen der Physik}\ }\textbf {\bibinfo {volume} {531}},\ \bibinfo
  {pages} {1900347} (\bibinfo {year} {2019})}\BibitemShut {NoStop}%
\bibitem [{\citenamefont {Cohen}\ \emph {et~al.}(2020)\citenamefont {Cohen},
  \citenamefont {J{\"o}rg}, \citenamefont {Lumer}, \citenamefont {Plotnik},
  \citenamefont {Waller}, \citenamefont {Schulz}, \citenamefont {von
  Freymann},\ and\ \citenamefont {Segev}}]{cohen2020generalized}%
  \BibitemOpen
  \bibfield  {author} {\bibinfo {author} {\bibfnamefont {M.-I.}\ \bibnamefont
  {Cohen}}, \bibinfo {author} {\bibfnamefont {C.}~\bibnamefont {J{\"o}rg}},
  \bibinfo {author} {\bibfnamefont {Y.}~\bibnamefont {Lumer}}, \bibinfo
  {author} {\bibfnamefont {Y.}~\bibnamefont {Plotnik}}, \bibinfo {author}
  {\bibfnamefont {E.~H.}\ \bibnamefont {Waller}}, \bibinfo {author}
  {\bibfnamefont {J.}~\bibnamefont {Schulz}}, \bibinfo {author} {\bibfnamefont
  {G.}~\bibnamefont {von Freymann}},\ and\ \bibinfo {author} {\bibfnamefont
  {M.}~\bibnamefont {Segev}},\ }\bibfield  {title} {\bibinfo {title}
  {Generalized laws of refraction and reflection at interfaces between
  different photonic artificial gauge fields},\ }\href@noop {} {\bibfield
  {journal} {\bibinfo  {journal} {Light: Science \& Applications}\ }\textbf
  {\bibinfo {volume} {9}},\ \bibinfo {pages} {1} (\bibinfo {year}
  {2020})}\BibitemShut {NoStop}%
\bibitem [{\citenamefont {Xue}\ \emph {et~al.}(2021)\citenamefont {Xue},
  \citenamefont {Yang},\ and\ \citenamefont {Zhang}}]{xue2021topological}%
  \BibitemOpen
  \bibfield  {author} {\bibinfo {author} {\bibfnamefont {H.}~\bibnamefont
  {Xue}}, \bibinfo {author} {\bibfnamefont {Y.}~\bibnamefont {Yang}},\ and\
  \bibinfo {author} {\bibfnamefont {B.}~\bibnamefont {Zhang}},\ }\bibfield
  {title} {\bibinfo {title} {Topological valley photonics: physics and device
  applications},\ }\href@noop {} {\bibfield  {journal} {\bibinfo  {journal}
  {Advanced Photonics Research}\ }\textbf {\bibinfo {volume} {2}},\ \bibinfo
  {pages} {2100013} (\bibinfo {year} {2021})}\BibitemShut {NoStop}%
\bibitem [{\citenamefont {Yang}\ \emph
  {et~al.}(2020{\natexlab{a}})\citenamefont {Yang}, \citenamefont {Yamagami},
  \citenamefont {Yu}, \citenamefont {Pitchappa}, \citenamefont {Webber},
  \citenamefont {Zhang}, \citenamefont {Fujita}, \citenamefont {Nagatsuma},\
  and\ \citenamefont {Singh}}]{yang2020terahertz}%
  \BibitemOpen
  \bibfield  {author} {\bibinfo {author} {\bibfnamefont {Y.}~\bibnamefont
  {Yang}}, \bibinfo {author} {\bibfnamefont {Y.}~\bibnamefont {Yamagami}},
  \bibinfo {author} {\bibfnamefont {X.}~\bibnamefont {Yu}}, \bibinfo {author}
  {\bibfnamefont {P.}~\bibnamefont {Pitchappa}}, \bibinfo {author}
  {\bibfnamefont {J.}~\bibnamefont {Webber}}, \bibinfo {author} {\bibfnamefont
  {B.}~\bibnamefont {Zhang}}, \bibinfo {author} {\bibfnamefont
  {M.}~\bibnamefont {Fujita}}, \bibinfo {author} {\bibfnamefont
  {T.}~\bibnamefont {Nagatsuma}},\ and\ \bibinfo {author} {\bibfnamefont
  {R.}~\bibnamefont {Singh}},\ }\bibfield  {title} {\bibinfo {title} {Terahertz
  topological photonics for on-chip communication},\ }\href@noop {} {\bibfield
  {journal} {\bibinfo  {journal} {Nature Photonics}\ }\textbf {\bibinfo
  {volume} {14}},\ \bibinfo {pages} {446} (\bibinfo {year}
  {2020}{\natexlab{a}})}\BibitemShut {NoStop}%
\bibitem [{\citenamefont {Nemoto}\ \emph {et~al.}(2014)\citenamefont {Nemoto},
  \citenamefont {Trupke}, \citenamefont {Devitt}, \citenamefont {Stephens},
  \citenamefont {Scharfenberger}, \citenamefont {Buczak}, \citenamefont
  {N{\"o}bauer}, \citenamefont {Everitt}, \citenamefont {Schmiedmayer},\ and\
  \citenamefont {Munro}}]{nemoto2014photonic}%
  \BibitemOpen
  \bibfield  {author} {\bibinfo {author} {\bibfnamefont {K.}~\bibnamefont
  {Nemoto}}, \bibinfo {author} {\bibfnamefont {M.}~\bibnamefont {Trupke}},
  \bibinfo {author} {\bibfnamefont {S.~J.}\ \bibnamefont {Devitt}}, \bibinfo
  {author} {\bibfnamefont {A.~M.}\ \bibnamefont {Stephens}}, \bibinfo {author}
  {\bibfnamefont {B.}~\bibnamefont {Scharfenberger}}, \bibinfo {author}
  {\bibfnamefont {K.}~\bibnamefont {Buczak}}, \bibinfo {author} {\bibfnamefont
  {T.}~\bibnamefont {N{\"o}bauer}}, \bibinfo {author} {\bibfnamefont {M.~S.}\
  \bibnamefont {Everitt}}, \bibinfo {author} {\bibfnamefont {J.}~\bibnamefont
  {Schmiedmayer}},\ and\ \bibinfo {author} {\bibfnamefont {W.~J.}\ \bibnamefont
  {Munro}},\ }\bibfield  {title} {\bibinfo {title} {Photonic architecture for
  scalable quantum information processing in diamond},\ }\href@noop {}
  {\bibfield  {journal} {\bibinfo  {journal} {Physical Review X}\ }\textbf
  {\bibinfo {volume} {4}},\ \bibinfo {pages} {031022} (\bibinfo {year}
  {2014})}\BibitemShut {NoStop}%
\bibitem [{\citenamefont {Chen}\ \emph {et~al.}(2021)\citenamefont {Chen},
  \citenamefont {He}, \citenamefont {Cheng}, \citenamefont {Qiu}, \citenamefont
  {Feng}, \citenamefont {Zhang}, \citenamefont {Dai}, \citenamefont {Guo},
  \citenamefont {Dong},\ and\ \citenamefont {Ren}}]{chen2021topologically}%
  \BibitemOpen
  \bibfield  {author} {\bibinfo {author} {\bibfnamefont {Y.}~\bibnamefont
  {Chen}}, \bibinfo {author} {\bibfnamefont {X.-T.}\ \bibnamefont {He}},
  \bibinfo {author} {\bibfnamefont {Y.-J.}\ \bibnamefont {Cheng}}, \bibinfo
  {author} {\bibfnamefont {H.-Y.}\ \bibnamefont {Qiu}}, \bibinfo {author}
  {\bibfnamefont {L.-T.}\ \bibnamefont {Feng}}, \bibinfo {author}
  {\bibfnamefont {M.}~\bibnamefont {Zhang}}, \bibinfo {author} {\bibfnamefont
  {D.-X.}\ \bibnamefont {Dai}}, \bibinfo {author} {\bibfnamefont {G.-C.}\
  \bibnamefont {Guo}}, \bibinfo {author} {\bibfnamefont {J.-W.}\ \bibnamefont
  {Dong}},\ and\ \bibinfo {author} {\bibfnamefont {X.-F.}\ \bibnamefont
  {Ren}},\ }\bibfield  {title} {\bibinfo {title} {Topologically protected
  valley-dependent quantum photonic circuits},\ }\href@noop {} {\bibfield
  {journal} {\bibinfo  {journal} {Physical Review Letters}\ }\textbf {\bibinfo
  {volume} {126}},\ \bibinfo {pages} {230503} (\bibinfo {year}
  {2021})}\BibitemShut {NoStop}%
\bibitem [{\citenamefont {St-Jean}\ \emph {et~al.}(2017)\citenamefont
  {St-Jean}, \citenamefont {Goblot}, \citenamefont {Galopin}, \citenamefont
  {Lema{\^\i}tre}, \citenamefont {Ozawa}, \citenamefont {Le~Gratiet},
  \citenamefont {Sagnes}, \citenamefont {Bloch},\ and\ \citenamefont
  {Amo}}]{st2017lasing}%
  \BibitemOpen
  \bibfield  {author} {\bibinfo {author} {\bibfnamefont {P.}~\bibnamefont
  {St-Jean}}, \bibinfo {author} {\bibfnamefont {V.}~\bibnamefont {Goblot}},
  \bibinfo {author} {\bibfnamefont {E.}~\bibnamefont {Galopin}}, \bibinfo
  {author} {\bibfnamefont {A.}~\bibnamefont {Lema{\^\i}tre}}, \bibinfo {author}
  {\bibfnamefont {T.}~\bibnamefont {Ozawa}}, \bibinfo {author} {\bibfnamefont
  {L.}~\bibnamefont {Le~Gratiet}}, \bibinfo {author} {\bibfnamefont
  {I.}~\bibnamefont {Sagnes}}, \bibinfo {author} {\bibfnamefont
  {J.}~\bibnamefont {Bloch}},\ and\ \bibinfo {author} {\bibfnamefont
  {A.}~\bibnamefont {Amo}},\ }\bibfield  {title} {\bibinfo {title} {Lasing in
  topological edge states of a one-dimensional lattice},\ }\href@noop {}
  {\bibfield  {journal} {\bibinfo  {journal} {Nature Photonics}\ }\textbf
  {\bibinfo {volume} {11}},\ \bibinfo {pages} {651} (\bibinfo {year}
  {2017})}\BibitemShut {NoStop}%
\bibitem [{\citenamefont {Xiao}\ \emph {et~al.}(2014)\citenamefont {Xiao},
  \citenamefont {Zhang},\ and\ \citenamefont {Chan}}]{xiao2014surface}%
  \BibitemOpen
  \bibfield  {author} {\bibinfo {author} {\bibfnamefont {M.}~\bibnamefont
  {Xiao}}, \bibinfo {author} {\bibfnamefont {Z.}~\bibnamefont {Zhang}},\ and\
  \bibinfo {author} {\bibfnamefont {C.~T.}\ \bibnamefont {Chan}},\ }\bibfield
  {title} {\bibinfo {title} {Surface impedance and bulk band geometric phases
  in one-dimensional systems},\ }\href@noop {} {\bibfield  {journal} {\bibinfo
  {journal} {Physical Review X}\ }\textbf {\bibinfo {volume} {4}},\ \bibinfo
  {pages} {021017} (\bibinfo {year} {2014})}\BibitemShut {NoStop}%
\bibitem [{\citenamefont {Hu}\ \emph {et~al.}(2021)\citenamefont {Hu},
  \citenamefont {Zhang}, \citenamefont {Jiang}, \citenamefont {Qiao},
  \citenamefont {Wang}, \citenamefont {Zhu}, \citenamefont {Xiao},\ and\
  \citenamefont {Liu}}]{hu2021double}%
  \BibitemOpen
  \bibfield  {author} {\bibinfo {author} {\bibfnamefont {M.}~\bibnamefont
  {Hu}}, \bibinfo {author} {\bibfnamefont {Y.}~\bibnamefont {Zhang}}, \bibinfo
  {author} {\bibfnamefont {X.}~\bibnamefont {Jiang}}, \bibinfo {author}
  {\bibfnamefont {T.}~\bibnamefont {Qiao}}, \bibinfo {author} {\bibfnamefont
  {Q.}~\bibnamefont {Wang}}, \bibinfo {author} {\bibfnamefont {S.}~\bibnamefont
  {Zhu}}, \bibinfo {author} {\bibfnamefont {M.}~\bibnamefont {Xiao}},\ and\
  \bibinfo {author} {\bibfnamefont {H.}~\bibnamefont {Liu}},\ }\bibfield
  {title} {\bibinfo {title} {Double-bowl state in photonic dirac nodal line
  semimetal},\ }\href@noop {} {\bibfield  {journal} {\bibinfo  {journal}
  {Light: Science \& Applications}\ }\textbf {\bibinfo {volume} {10}},\
  \bibinfo {pages} {1} (\bibinfo {year} {2021})}\BibitemShut {NoStop}%
\bibitem [{\citenamefont {Tan}(2021)}]{tan2021topological}%
  \BibitemOpen
  \bibfield  {author} {\bibinfo {author} {\bibfnamefont {D.~T.}\ \bibnamefont
  {Tan}},\ }\bibfield  {title} {\bibinfo {title} {Topological silicon
  photonics},\ }\href@noop {} {\bibfield  {journal} {\bibinfo  {journal}
  {Advanced Photonics Research}\ }\textbf {\bibinfo {volume} {2}},\ \bibinfo
  {pages} {2100010} (\bibinfo {year} {2021})}\BibitemShut {NoStop}%
\bibitem [{\citenamefont {Lin}\ \emph {et~al.}(2021)\citenamefont {Lin},
  \citenamefont {Chen},\ and\ \citenamefont {Hsueh}}]{lin2021conjugated}%
  \BibitemOpen
  \bibfield  {author} {\bibinfo {author} {\bibfnamefont {Y.-C.}\ \bibnamefont
  {Lin}}, \bibinfo {author} {\bibfnamefont {B.-Y.}\ \bibnamefont {Chen}},\ and\
  \bibinfo {author} {\bibfnamefont {W.-J.}\ \bibnamefont {Hsueh}},\ }\bibfield
  {title} {\bibinfo {title} {Conjugated topological interface-states in coupled
  ring resonators},\ }\href@noop {} {\bibfield  {journal} {\bibinfo  {journal}
  {Scientific Reports}\ }\textbf {\bibinfo {volume} {11}},\ \bibinfo {pages}
  {1} (\bibinfo {year} {2021})}\BibitemShut {NoStop}%
\bibitem [{\citenamefont {Hu}\ \emph {et~al.}(2019)\citenamefont {Hu},
  \citenamefont {Liu}, \citenamefont {Xie}, \citenamefont {Zhang},
  \citenamefont {Yao}, \citenamefont {Zhang},\ and\ \citenamefont
  {Zhan}}]{hu2019strong}%
  \BibitemOpen
  \bibfield  {author} {\bibinfo {author} {\bibfnamefont {J.}~\bibnamefont
  {Hu}}, \bibinfo {author} {\bibfnamefont {W.}~\bibnamefont {Liu}}, \bibinfo
  {author} {\bibfnamefont {W.}~\bibnamefont {Xie}}, \bibinfo {author}
  {\bibfnamefont {W.}~\bibnamefont {Zhang}}, \bibinfo {author} {\bibfnamefont
  {E.}~\bibnamefont {Yao}}, \bibinfo {author} {\bibfnamefont {Y.}~\bibnamefont
  {Zhang}},\ and\ \bibinfo {author} {\bibfnamefont {Q.}~\bibnamefont {Zhan}},\
  }\bibfield  {title} {\bibinfo {title} {Strong coupling of optical interface
  modes in a 1d topological photonic crystal heterostructure/ag hybrid
  system},\ }\href@noop {} {\bibfield  {journal} {\bibinfo  {journal} {Optics
  Letters}\ }\textbf {\bibinfo {volume} {44}},\ \bibinfo {pages} {5642}
  (\bibinfo {year} {2019})}\BibitemShut {NoStop}%
\bibitem [{\citenamefont {Gupta}\ \emph
  {et~al.}(2022{\natexlab{a}})\citenamefont {Gupta}, \citenamefont {Srinivasu},
  \citenamefont {Tiwari}, \citenamefont {Wanare},\ and\ \citenamefont
  {Ramakrishna}}]{gupta2022realizing}%
  \BibitemOpen
  \bibfield  {author} {\bibinfo {author} {\bibfnamefont {N.~K.}\ \bibnamefont
  {Gupta}}, \bibinfo {author} {\bibfnamefont {S.}~\bibnamefont {Srinivasu}},
  \bibinfo {author} {\bibfnamefont {A.~K.}\ \bibnamefont {Tiwari}}, \bibinfo
  {author} {\bibfnamefont {H.}~\bibnamefont {Wanare}},\ and\ \bibinfo {author}
  {\bibfnamefont {S.~A.}\ \bibnamefont {Ramakrishna}},\ }\bibfield  {title}
  {\bibinfo {title} {Realizing quasi-monochromatic switchable thermal emission
  from electro-optically induced topological phase transitions},\ }\href@noop
  {} {\bibfield  {journal} {\bibinfo  {journal} {Scientific Reports}\ }\textbf
  {\bibinfo {volume} {12}},\ \bibinfo {pages} {1} (\bibinfo {year}
  {2022}{\natexlab{a}})}\BibitemShut {NoStop}%
\bibitem [{\citenamefont {Khanikaev}\ \emph {et~al.}(2013)\citenamefont
  {Khanikaev}, \citenamefont {Hossein~Mousavi}, \citenamefont {Tse},
  \citenamefont {Kargarian}, \citenamefont {MacDonald},\ and\ \citenamefont
  {Shvets}}]{khanikaev2013photonic}%
  \BibitemOpen
  \bibfield  {author} {\bibinfo {author} {\bibfnamefont {A.~B.}\ \bibnamefont
  {Khanikaev}}, \bibinfo {author} {\bibfnamefont {S.}~\bibnamefont
  {Hossein~Mousavi}}, \bibinfo {author} {\bibfnamefont {W.-K.}\ \bibnamefont
  {Tse}}, \bibinfo {author} {\bibfnamefont {M.}~\bibnamefont {Kargarian}},
  \bibinfo {author} {\bibfnamefont {A.~H.}\ \bibnamefont {MacDonald}},\ and\
  \bibinfo {author} {\bibfnamefont {G.}~\bibnamefont {Shvets}},\ }\bibfield
  {title} {\bibinfo {title} {Photonic topological insulators},\ }\href@noop {}
  {\bibfield  {journal} {\bibinfo  {journal} {Nature Materials}\ }\textbf
  {\bibinfo {volume} {12}},\ \bibinfo {pages} {233} (\bibinfo {year}
  {2013})}\BibitemShut {NoStop}%
\bibitem [{\citenamefont {Yang}\ \emph {et~al.}(2019)\citenamefont {Yang},
  \citenamefont {Gao}, \citenamefont {Xue}, \citenamefont {Zhang},
  \citenamefont {He}, \citenamefont {Yang}, \citenamefont {Singh},
  \citenamefont {Chong}, \citenamefont {Zhang},\ and\ \citenamefont
  {Chen}}]{yang2019realization}%
  \BibitemOpen
  \bibfield  {author} {\bibinfo {author} {\bibfnamefont {Y.}~\bibnamefont
  {Yang}}, \bibinfo {author} {\bibfnamefont {Z.}~\bibnamefont {Gao}}, \bibinfo
  {author} {\bibfnamefont {H.}~\bibnamefont {Xue}}, \bibinfo {author}
  {\bibfnamefont {L.}~\bibnamefont {Zhang}}, \bibinfo {author} {\bibfnamefont
  {M.}~\bibnamefont {He}}, \bibinfo {author} {\bibfnamefont {Z.}~\bibnamefont
  {Yang}}, \bibinfo {author} {\bibfnamefont {R.}~\bibnamefont {Singh}},
  \bibinfo {author} {\bibfnamefont {Y.}~\bibnamefont {Chong}}, \bibinfo
  {author} {\bibfnamefont {B.}~\bibnamefont {Zhang}},\ and\ \bibinfo {author}
  {\bibfnamefont {H.}~\bibnamefont {Chen}},\ }\bibfield  {title} {\bibinfo
  {title} {Realization of a three-dimensional photonic topological insulator},\
  }\href@noop {} {\bibfield  {journal} {\bibinfo  {journal} {Nature}\ }\textbf
  {\bibinfo {volume} {565}},\ \bibinfo {pages} {622} (\bibinfo {year}
  {2019})}\BibitemShut {NoStop}%
\bibitem [{\citenamefont {Lustig}\ \emph {et~al.}(2019)\citenamefont {Lustig},
  \citenamefont {Weimann}, \citenamefont {Plotnik}, \citenamefont {Lumer},
  \citenamefont {Bandres}, \citenamefont {Szameit},\ and\ \citenamefont
  {Segev}}]{lustig2019photonic}%
  \BibitemOpen
  \bibfield  {author} {\bibinfo {author} {\bibfnamefont {E.}~\bibnamefont
  {Lustig}}, \bibinfo {author} {\bibfnamefont {S.}~\bibnamefont {Weimann}},
  \bibinfo {author} {\bibfnamefont {Y.}~\bibnamefont {Plotnik}}, \bibinfo
  {author} {\bibfnamefont {Y.}~\bibnamefont {Lumer}}, \bibinfo {author}
  {\bibfnamefont {M.~A.}\ \bibnamefont {Bandres}}, \bibinfo {author}
  {\bibfnamefont {A.}~\bibnamefont {Szameit}},\ and\ \bibinfo {author}
  {\bibfnamefont {M.}~\bibnamefont {Segev}},\ }\bibfield  {title} {\bibinfo
  {title} {Photonic topological insulator in synthetic dimensions},\
  }\href@noop {} {\bibfield  {journal} {\bibinfo  {journal} {Nature}\ }\textbf
  {\bibinfo {volume} {567}},\ \bibinfo {pages} {356} (\bibinfo {year}
  {2019})}\BibitemShut {NoStop}%
\bibitem [{\citenamefont {Kim}\ \emph {et~al.}(2022)\citenamefont {Kim},
  \citenamefont {Wang}, \citenamefont {Yang}, \citenamefont {Teo},
  \citenamefont {Rho},\ and\ \citenamefont {Zhang}}]{kim2022three}%
  \BibitemOpen
  \bibfield  {author} {\bibinfo {author} {\bibfnamefont {M.}~\bibnamefont
  {Kim}}, \bibinfo {author} {\bibfnamefont {Z.}~\bibnamefont {Wang}}, \bibinfo
  {author} {\bibfnamefont {Y.}~\bibnamefont {Yang}}, \bibinfo {author}
  {\bibfnamefont {H.~T.}\ \bibnamefont {Teo}}, \bibinfo {author} {\bibfnamefont
  {J.}~\bibnamefont {Rho}},\ and\ \bibinfo {author} {\bibfnamefont
  {B.}~\bibnamefont {Zhang}},\ }\bibfield  {title} {\bibinfo {title}
  {Three-dimensional photonic topological insulator without spin--orbit
  coupling},\ }\href@noop {} {\bibfield  {journal} {\bibinfo  {journal} {Nature
  Communications}\ }\textbf {\bibinfo {volume} {13}},\ \bibinfo {pages} {1}
  (\bibinfo {year} {2022})}\BibitemShut {NoStop}%
\bibitem [{\citenamefont {Tan}\ \emph {et~al.}(2022)\citenamefont {Tan},
  \citenamefont {Wang}, \citenamefont {Kumar},\ and\ \citenamefont
  {Singh}}]{tan2022interfacial}%
  \BibitemOpen
  \bibfield  {author} {\bibinfo {author} {\bibfnamefont {Y.~J.}\ \bibnamefont
  {Tan}}, \bibinfo {author} {\bibfnamefont {W.}~\bibnamefont {Wang}}, \bibinfo
  {author} {\bibfnamefont {A.}~\bibnamefont {Kumar}},\ and\ \bibinfo {author}
  {\bibfnamefont {R.}~\bibnamefont {Singh}},\ }\bibfield  {title} {\bibinfo
  {title} {Interfacial topological photonics: broadband silicon waveguides for
  thz 6g communication and beyond},\ }\href@noop {} {\bibfield  {journal}
  {\bibinfo  {journal} {Optics Express}\ }\textbf {\bibinfo {volume} {30}},\
  \bibinfo {pages} {33035} (\bibinfo {year} {2022})}\BibitemShut {NoStop}%
\bibitem [{\citenamefont {Liu}\ \emph {et~al.}(2022)\citenamefont {Liu},
  \citenamefont {Kobayashi}, \citenamefont {Ikeda}, \citenamefont {Ota},\ and\
  \citenamefont {Iwamoto}}]{liu2022topological}%
  \BibitemOpen
  \bibfield  {author} {\bibinfo {author} {\bibfnamefont {T.}~\bibnamefont
  {Liu}}, \bibinfo {author} {\bibfnamefont {N.}~\bibnamefont {Kobayashi}},
  \bibinfo {author} {\bibfnamefont {K.}~\bibnamefont {Ikeda}}, \bibinfo
  {author} {\bibfnamefont {Y.}~\bibnamefont {Ota}},\ and\ \bibinfo {author}
  {\bibfnamefont {S.}~\bibnamefont {Iwamoto}},\ }\bibfield  {title} {\bibinfo
  {title} {Topological band gaps enlarged in epsilon-near-zero magneto-optical
  photonic crystals},\ }\href@noop {} {\bibfield  {journal} {\bibinfo
  {journal} {ACS Photonics}\ } (\bibinfo {year} {2022})}\BibitemShut {NoStop}%
\bibitem [{\citenamefont {Chen}\ \emph {et~al.}(2014)\citenamefont {Chen},
  \citenamefont {Jiang}, \citenamefont {Chen}, \citenamefont {Zhu},
  \citenamefont {Zhou}, \citenamefont {Dong},\ and\ \citenamefont
  {Chan}}]{chen2014experimental}%
  \BibitemOpen
  \bibfield  {author} {\bibinfo {author} {\bibfnamefont {W.-J.}\ \bibnamefont
  {Chen}}, \bibinfo {author} {\bibfnamefont {S.-J.}\ \bibnamefont {Jiang}},
  \bibinfo {author} {\bibfnamefont {X.-D.}\ \bibnamefont {Chen}}, \bibinfo
  {author} {\bibfnamefont {B.}~\bibnamefont {Zhu}}, \bibinfo {author}
  {\bibfnamefont {L.}~\bibnamefont {Zhou}}, \bibinfo {author} {\bibfnamefont
  {J.-W.}\ \bibnamefont {Dong}},\ and\ \bibinfo {author} {\bibfnamefont
  {C.~T.}\ \bibnamefont {Chan}},\ }\bibfield  {title} {\bibinfo {title}
  {Experimental realization of photonic topological insulator in a uniaxial
  metacrystal waveguide},\ }\href@noop {} {\bibfield  {journal} {\bibinfo
  {journal} {Nature Communications}\ }\textbf {\bibinfo {volume} {5}},\
  \bibinfo {pages} {1} (\bibinfo {year} {2014})}\BibitemShut {NoStop}%
\bibitem [{\citenamefont {Kumar}\ \emph {et~al.}(2022)\citenamefont {Kumar},
  \citenamefont {Gupta}, \citenamefont {Pitchappa}, \citenamefont {Tan},
  \citenamefont {Wang},\ and\ \citenamefont {Singh}}]{kumar2022topological}%
  \BibitemOpen
  \bibfield  {author} {\bibinfo {author} {\bibfnamefont {A.}~\bibnamefont
  {Kumar}}, \bibinfo {author} {\bibfnamefont {M.}~\bibnamefont {Gupta}},
  \bibinfo {author} {\bibfnamefont {P.}~\bibnamefont {Pitchappa}}, \bibinfo
  {author} {\bibfnamefont {Y.~J.}\ \bibnamefont {Tan}}, \bibinfo {author}
  {\bibfnamefont {N.}~\bibnamefont {Wang}},\ and\ \bibinfo {author}
  {\bibfnamefont {R.}~\bibnamefont {Singh}},\ }\bibfield  {title} {\bibinfo
  {title} {Topological sensor on a silicon chip},\ }\href@noop {} {\bibfield
  {journal} {\bibinfo  {journal} {Applied Physics Letters}\ }\textbf {\bibinfo
  {volume} {121}},\ \bibinfo {pages} {011101} (\bibinfo {year}
  {2022})}\BibitemShut {NoStop}%
\bibitem [{\citenamefont {Kurganov}\ \emph {et~al.}(2022)\citenamefont
  {Kurganov}, \citenamefont {Dobrykh}, \citenamefont {Puhtina}, \citenamefont
  {Yusupov}, \citenamefont {Slobozhanyuk}, \citenamefont {Kivshar},\ and\
  \citenamefont {Zhirihin}}]{kurganov2022temperature}%
  \BibitemOpen
  \bibfield  {author} {\bibinfo {author} {\bibfnamefont {G.}~\bibnamefont
  {Kurganov}}, \bibinfo {author} {\bibfnamefont {D.}~\bibnamefont {Dobrykh}},
  \bibinfo {author} {\bibfnamefont {E.}~\bibnamefont {Puhtina}}, \bibinfo
  {author} {\bibfnamefont {I.}~\bibnamefont {Yusupov}}, \bibinfo {author}
  {\bibfnamefont {A.}~\bibnamefont {Slobozhanyuk}}, \bibinfo {author}
  {\bibfnamefont {Y.~S.}\ \bibnamefont {Kivshar}},\ and\ \bibinfo {author}
  {\bibfnamefont {D.}~\bibnamefont {Zhirihin}},\ }\bibfield  {title} {\bibinfo
  {title} {Temperature control of electromagnetic topological edge states},\
  }\href@noop {} {\bibfield  {journal} {\bibinfo  {journal} {Applied Physics
  Letters}\ }\textbf {\bibinfo {volume} {120}},\ \bibinfo {pages} {233105}
  (\bibinfo {year} {2022})}\BibitemShut {NoStop}%
\bibitem [{\citenamefont {Yang}\ \emph
  {et~al.}(2020{\natexlab{b}})\citenamefont {Yang}, \citenamefont {Lustig},
  \citenamefont {Lumer},\ and\ \citenamefont {Segev}}]{yang2020photonic}%
  \BibitemOpen
  \bibfield  {author} {\bibinfo {author} {\bibfnamefont {Z.}~\bibnamefont
  {Yang}}, \bibinfo {author} {\bibfnamefont {E.}~\bibnamefont {Lustig}},
  \bibinfo {author} {\bibfnamefont {Y.}~\bibnamefont {Lumer}},\ and\ \bibinfo
  {author} {\bibfnamefont {M.}~\bibnamefont {Segev}},\ }\bibfield  {title}
  {\bibinfo {title} {Photonic floquet topological insulators in a fractal
  lattice},\ }\href@noop {} {\bibfield  {journal} {\bibinfo  {journal} {Light:
  Science \& Applications}\ }\textbf {\bibinfo {volume} {9}},\ \bibinfo {pages}
  {1} (\bibinfo {year} {2020}{\natexlab{b}})}\BibitemShut {NoStop}%
\bibitem [{\citenamefont {Pocock}\ \emph {et~al.}(2018)\citenamefont {Pocock},
  \citenamefont {Xiao}, \citenamefont {Huidobro},\ and\ \citenamefont
  {Giannini}}]{pocock2018topological}%
  \BibitemOpen
  \bibfield  {author} {\bibinfo {author} {\bibfnamefont {S.~R.}\ \bibnamefont
  {Pocock}}, \bibinfo {author} {\bibfnamefont {X.}~\bibnamefont {Xiao}},
  \bibinfo {author} {\bibfnamefont {P.~A.}\ \bibnamefont {Huidobro}},\ and\
  \bibinfo {author} {\bibfnamefont {V.}~\bibnamefont {Giannini}},\ }\bibfield
  {title} {\bibinfo {title} {Topological plasmonic chain with retardation and
  radiative effects},\ }\href@noop {} {\bibfield  {journal} {\bibinfo
  {journal} {ACS Photonics}\ }\textbf {\bibinfo {volume} {5}},\ \bibinfo
  {pages} {2271} (\bibinfo {year} {2018})}\BibitemShut {NoStop}%
\bibitem [{\citenamefont {St-Jean}\ \emph {et~al.}(2021)\citenamefont
  {St-Jean}, \citenamefont {Dauphin}, \citenamefont {Massignan}, \citenamefont
  {Real}, \citenamefont {Jamadi}, \citenamefont {Milicevic}, \citenamefont
  {Lemaitre}, \citenamefont {Harouri}, \citenamefont {Le~Gratiet},
  \citenamefont {Sagnes} \emph {et~al.}}]{st2021measuring}%
  \BibitemOpen
  \bibfield  {author} {\bibinfo {author} {\bibfnamefont {P.}~\bibnamefont
  {St-Jean}}, \bibinfo {author} {\bibfnamefont {A.}~\bibnamefont {Dauphin}},
  \bibinfo {author} {\bibfnamefont {P.}~\bibnamefont {Massignan}}, \bibinfo
  {author} {\bibfnamefont {B.}~\bibnamefont {Real}}, \bibinfo {author}
  {\bibfnamefont {O.}~\bibnamefont {Jamadi}}, \bibinfo {author} {\bibfnamefont
  {M.}~\bibnamefont {Milicevic}}, \bibinfo {author} {\bibfnamefont
  {A.}~\bibnamefont {Lemaitre}}, \bibinfo {author} {\bibfnamefont
  {A.}~\bibnamefont {Harouri}}, \bibinfo {author} {\bibfnamefont
  {L.}~\bibnamefont {Le~Gratiet}}, \bibinfo {author} {\bibfnamefont
  {I.}~\bibnamefont {Sagnes}}, \emph {et~al.},\ }\bibfield  {title} {\bibinfo
  {title} {Measuring topological invariants in a polaritonic analog of
  graphene},\ }\href@noop {} {\bibfield  {journal} {\bibinfo  {journal}
  {Physical Review Letters}\ }\textbf {\bibinfo {volume} {126}},\ \bibinfo
  {pages} {127403} (\bibinfo {year} {2021})}\BibitemShut {NoStop}%
\bibitem [{\citenamefont {Rappoport}\ \emph {et~al.}(2021)\citenamefont
  {Rappoport}, \citenamefont {Bludov}, \citenamefont {Koppens},\ and\
  \citenamefont {Peres}}]{rappoport2021topological}%
  \BibitemOpen
  \bibfield  {author} {\bibinfo {author} {\bibfnamefont {T.~G.}\ \bibnamefont
  {Rappoport}}, \bibinfo {author} {\bibfnamefont {Y.~V.}\ \bibnamefont
  {Bludov}}, \bibinfo {author} {\bibfnamefont {F.~H.}\ \bibnamefont
  {Koppens}},\ and\ \bibinfo {author} {\bibfnamefont {N.~M.}\ \bibnamefont
  {Peres}},\ }\bibfield  {title} {\bibinfo {title} {Topological graphene
  plasmons in a plasmonic realization of the su--schrieffer--heeger model},\
  }\href@noop {} {\bibfield  {journal} {\bibinfo  {journal} {ACS Photonics}\
  }\textbf {\bibinfo {volume} {8}},\ \bibinfo {pages} {1817} (\bibinfo {year}
  {2021})}\BibitemShut {NoStop}%
\bibitem [{\citenamefont {Liu}\ \emph {et~al.}(2021)\citenamefont {Liu},
  \citenamefont {Qin}, \citenamefont {Liu}, \citenamefont {Zheng},
  \citenamefont {Ren}, \citenamefont {Wang},\ and\ \citenamefont
  {Lu}}]{liu2021frequency}%
  \BibitemOpen
  \bibfield  {author} {\bibinfo {author} {\bibfnamefont {Z.}~\bibnamefont
  {Liu}}, \bibinfo {author} {\bibfnamefont {C.}~\bibnamefont {Qin}}, \bibinfo
  {author} {\bibfnamefont {W.}~\bibnamefont {Liu}}, \bibinfo {author}
  {\bibfnamefont {L.}~\bibnamefont {Zheng}}, \bibinfo {author} {\bibfnamefont
  {S.}~\bibnamefont {Ren}}, \bibinfo {author} {\bibfnamefont {B.}~\bibnamefont
  {Wang}},\ and\ \bibinfo {author} {\bibfnamefont {P.}~\bibnamefont {Lu}},\
  }\bibfield  {title} {\bibinfo {title} {Frequency manipulation of topological
  surface states by weyl phase transitions},\ }\href@noop {} {\bibfield
  {journal} {\bibinfo  {journal} {Optics Letters}\ }\textbf {\bibinfo {volume}
  {46}},\ \bibinfo {pages} {5719} (\bibinfo {year} {2021})}\BibitemShut
  {NoStop}%
\bibitem [{\citenamefont {Xia}\ \emph {et~al.}(2020)\citenamefont {Xia},
  \citenamefont {Juki{\'c}}, \citenamefont {Wang}, \citenamefont {Smirnova},
  \citenamefont {Smirnov}, \citenamefont {Tang}, \citenamefont {Song},
  \citenamefont {Szameit}, \citenamefont {Leykam}, \citenamefont {Xu} \emph
  {et~al.}}]{xia2020nontrivial}%
  \BibitemOpen
  \bibfield  {author} {\bibinfo {author} {\bibfnamefont {S.}~\bibnamefont
  {Xia}}, \bibinfo {author} {\bibfnamefont {D.}~\bibnamefont {Juki{\'c}}},
  \bibinfo {author} {\bibfnamefont {N.}~\bibnamefont {Wang}}, \bibinfo {author}
  {\bibfnamefont {D.}~\bibnamefont {Smirnova}}, \bibinfo {author}
  {\bibfnamefont {L.}~\bibnamefont {Smirnov}}, \bibinfo {author} {\bibfnamefont
  {L.}~\bibnamefont {Tang}}, \bibinfo {author} {\bibfnamefont {D.}~\bibnamefont
  {Song}}, \bibinfo {author} {\bibfnamefont {A.}~\bibnamefont {Szameit}},
  \bibinfo {author} {\bibfnamefont {D.}~\bibnamefont {Leykam}}, \bibinfo
  {author} {\bibfnamefont {J.}~\bibnamefont {Xu}}, \emph {et~al.},\ }\bibfield
  {title} {\bibinfo {title} {Nontrivial coupling of light into a defect: the
  interplay of nonlinearity and topology},\ }\href@noop {} {\bibfield
  {journal} {\bibinfo  {journal} {Light: Science \& Applications}\ }\textbf
  {\bibinfo {volume} {9}},\ \bibinfo {pages} {1} (\bibinfo {year}
  {2020})}\BibitemShut {NoStop}%
\bibitem [{\citenamefont {Zak}(1989)}]{zak1989berry}%
  \BibitemOpen
  \bibfield  {author} {\bibinfo {author} {\bibfnamefont {J.}~\bibnamefont
  {Zak}},\ }\bibfield  {title} {\bibinfo {title} {Berry’s phase for energy
  bands in solids},\ }\href@noop {} {\bibfield  {journal} {\bibinfo  {journal}
  {Physical Review Letters}\ }\textbf {\bibinfo {volume} {62}},\ \bibinfo
  {pages} {2747} (\bibinfo {year} {1989})}\BibitemShut {NoStop}%
\bibitem [{\citenamefont {Atala}\ \emph {et~al.}(2013)\citenamefont {Atala},
  \citenamefont {Aidelsburger}, \citenamefont {Barreiro}, \citenamefont
  {Abanin}, \citenamefont {Kitagawa}, \citenamefont {Demler},\ and\
  \citenamefont {Bloch}}]{atala2013direct}%
  \BibitemOpen
  \bibfield  {author} {\bibinfo {author} {\bibfnamefont {M.}~\bibnamefont
  {Atala}}, \bibinfo {author} {\bibfnamefont {M.}~\bibnamefont {Aidelsburger}},
  \bibinfo {author} {\bibfnamefont {J.~T.}\ \bibnamefont {Barreiro}}, \bibinfo
  {author} {\bibfnamefont {D.}~\bibnamefont {Abanin}}, \bibinfo {author}
  {\bibfnamefont {T.}~\bibnamefont {Kitagawa}}, \bibinfo {author}
  {\bibfnamefont {E.}~\bibnamefont {Demler}},\ and\ \bibinfo {author}
  {\bibfnamefont {I.}~\bibnamefont {Bloch}},\ }\bibfield  {title} {\bibinfo
  {title} {Direct measurement of the zak phase in topological bloch bands},\
  }\href@noop {} {\bibfield  {journal} {\bibinfo  {journal} {Nature Physics}\
  }\textbf {\bibinfo {volume} {9}},\ \bibinfo {pages} {795} (\bibinfo {year}
  {2013})}\BibitemShut {NoStop}%
\bibitem [{\citenamefont {Jiao}\ \emph {et~al.}(2021)\citenamefont {Jiao},
  \citenamefont {Longhi}, \citenamefont {Wang}, \citenamefont {Gao},
  \citenamefont {Zhou}, \citenamefont {Wang}, \citenamefont {Fu}, \citenamefont
  {Wang}, \citenamefont {Ren}, \citenamefont {Qiao} \emph
  {et~al.}}]{jiao2021experimentally}%
  \BibitemOpen
  \bibfield  {author} {\bibinfo {author} {\bibfnamefont {Z.-Q.}\ \bibnamefont
  {Jiao}}, \bibinfo {author} {\bibfnamefont {S.}~\bibnamefont {Longhi}},
  \bibinfo {author} {\bibfnamefont {X.-W.}\ \bibnamefont {Wang}}, \bibinfo
  {author} {\bibfnamefont {J.}~\bibnamefont {Gao}}, \bibinfo {author}
  {\bibfnamefont {W.-H.}\ \bibnamefont {Zhou}}, \bibinfo {author}
  {\bibfnamefont {Y.}~\bibnamefont {Wang}}, \bibinfo {author} {\bibfnamefont
  {Y.-X.}\ \bibnamefont {Fu}}, \bibinfo {author} {\bibfnamefont
  {L.}~\bibnamefont {Wang}}, \bibinfo {author} {\bibfnamefont {R.-J.}\
  \bibnamefont {Ren}}, \bibinfo {author} {\bibfnamefont {L.-F.}\ \bibnamefont
  {Qiao}}, \emph {et~al.},\ }\bibfield  {title} {\bibinfo {title}
  {Experimentally detecting quantized zak phases without chiral symmetry in
  photonic lattices},\ }\href@noop {} {\bibfield  {journal} {\bibinfo
  {journal} {Physical Review Letters}\ }\textbf {\bibinfo {volume} {127}},\
  \bibinfo {pages} {147401} (\bibinfo {year} {2021})}\BibitemShut {NoStop}%
\bibitem [{\citenamefont {Abanin}\ \emph {et~al.}(2013)\citenamefont {Abanin},
  \citenamefont {Kitagawa}, \citenamefont {Bloch},\ and\ \citenamefont
  {Demler}}]{abanin2013interferometric}%
  \BibitemOpen
  \bibfield  {author} {\bibinfo {author} {\bibfnamefont {D.~A.}\ \bibnamefont
  {Abanin}}, \bibinfo {author} {\bibfnamefont {T.}~\bibnamefont {Kitagawa}},
  \bibinfo {author} {\bibfnamefont {I.}~\bibnamefont {Bloch}},\ and\ \bibinfo
  {author} {\bibfnamefont {E.}~\bibnamefont {Demler}},\ }\bibfield  {title}
  {\bibinfo {title} {Interferometric approach to measuring band topology in 2d
  optical lattices},\ }\href@noop {} {\bibfield  {journal} {\bibinfo  {journal}
  {Physical Review Letters}\ }\textbf {\bibinfo {volume} {110}},\ \bibinfo
  {pages} {165304} (\bibinfo {year} {2013})}\BibitemShut {NoStop}%
\bibitem [{\citenamefont {Mittal}\ \emph {et~al.}(2016)\citenamefont {Mittal},
  \citenamefont {Ganeshan}, \citenamefont {Fan}, \citenamefont {Vaezi},\ and\
  \citenamefont {Hafezi}}]{mittal2016measurement}%
  \BibitemOpen
  \bibfield  {author} {\bibinfo {author} {\bibfnamefont {S.}~\bibnamefont
  {Mittal}}, \bibinfo {author} {\bibfnamefont {S.}~\bibnamefont {Ganeshan}},
  \bibinfo {author} {\bibfnamefont {J.}~\bibnamefont {Fan}}, \bibinfo {author}
  {\bibfnamefont {A.}~\bibnamefont {Vaezi}},\ and\ \bibinfo {author}
  {\bibfnamefont {M.}~\bibnamefont {Hafezi}},\ }\bibfield  {title} {\bibinfo
  {title} {Measurement of topological invariants in a 2d photonic system},\
  }\href@noop {} {\bibfield  {journal} {\bibinfo  {journal} {Nature Photonics}\
  }\textbf {\bibinfo {volume} {10}},\ \bibinfo {pages} {180} (\bibinfo {year}
  {2016})}\BibitemShut {NoStop}%
\bibitem [{\citenamefont {Duca}\ \emph {et~al.}(2015)\citenamefont {Duca},
  \citenamefont {Li}, \citenamefont {Reitter}, \citenamefont {Bloch},
  \citenamefont {Schleier-Smith},\ and\ \citenamefont
  {Schneider}}]{duca2015aharonov}%
  \BibitemOpen
  \bibfield  {author} {\bibinfo {author} {\bibfnamefont {L.}~\bibnamefont
  {Duca}}, \bibinfo {author} {\bibfnamefont {T.}~\bibnamefont {Li}}, \bibinfo
  {author} {\bibfnamefont {M.}~\bibnamefont {Reitter}}, \bibinfo {author}
  {\bibfnamefont {I.}~\bibnamefont {Bloch}}, \bibinfo {author} {\bibfnamefont
  {M.}~\bibnamefont {Schleier-Smith}},\ and\ \bibinfo {author} {\bibfnamefont
  {U.}~\bibnamefont {Schneider}},\ }\bibfield  {title} {\bibinfo {title} {An
  aharonov-bohm interferometer for determining bloch band topology},\
  }\href@noop {} {\bibfield  {journal} {\bibinfo  {journal} {Science}\ }\textbf
  {\bibinfo {volume} {347}},\ \bibinfo {pages} {288} (\bibinfo {year}
  {2015})}\BibitemShut {NoStop}%
\bibitem [{\citenamefont {Flurin}\ \emph {et~al.}(2017)\citenamefont {Flurin},
  \citenamefont {Ramasesh}, \citenamefont {Hacohen-Gourgy}, \citenamefont
  {Martin}, \citenamefont {Yao},\ and\ \citenamefont
  {Siddiqi}}]{flurin2017observing}%
  \BibitemOpen
  \bibfield  {author} {\bibinfo {author} {\bibfnamefont {E.}~\bibnamefont
  {Flurin}}, \bibinfo {author} {\bibfnamefont {V.~V.}\ \bibnamefont
  {Ramasesh}}, \bibinfo {author} {\bibfnamefont {S.}~\bibnamefont
  {Hacohen-Gourgy}}, \bibinfo {author} {\bibfnamefont {L.~S.}\ \bibnamefont
  {Martin}}, \bibinfo {author} {\bibfnamefont {N.~Y.}\ \bibnamefont {Yao}},\
  and\ \bibinfo {author} {\bibfnamefont {I.}~\bibnamefont {Siddiqi}},\
  }\bibfield  {title} {\bibinfo {title} {Observing topological invariants using
  quantum walks in superconducting circuits},\ }\href@noop {} {\bibfield
  {journal} {\bibinfo  {journal} {Physical Review X}\ }\textbf {\bibinfo
  {volume} {7}},\ \bibinfo {pages} {031023} (\bibinfo {year}
  {2017})}\BibitemShut {NoStop}%
\bibitem [{\citenamefont {Ramasesh}\ \emph {et~al.}(2017)\citenamefont
  {Ramasesh}, \citenamefont {Flurin}, \citenamefont {Rudner}, \citenamefont
  {Siddiqi},\ and\ \citenamefont {Yao}}]{ramasesh2017direct}%
  \BibitemOpen
  \bibfield  {author} {\bibinfo {author} {\bibfnamefont {V.~V.}\ \bibnamefont
  {Ramasesh}}, \bibinfo {author} {\bibfnamefont {E.}~\bibnamefont {Flurin}},
  \bibinfo {author} {\bibfnamefont {M.}~\bibnamefont {Rudner}}, \bibinfo
  {author} {\bibfnamefont {I.}~\bibnamefont {Siddiqi}},\ and\ \bibinfo {author}
  {\bibfnamefont {N.~Y.}\ \bibnamefont {Yao}},\ }\bibfield  {title} {\bibinfo
  {title} {Direct probe of topological invariants using bloch oscillating
  quantum walks},\ }\href@noop {} {\bibfield  {journal} {\bibinfo  {journal}
  {Physical Review Letters}\ }\textbf {\bibinfo {volume} {118}},\ \bibinfo
  {pages} {130501} (\bibinfo {year} {2017})}\BibitemShut {NoStop}%
\bibitem [{\citenamefont {von Gersdorff}\ and\ \citenamefont
  {Chen}(2021)}]{von2021measurement}%
  \BibitemOpen
  \bibfield  {author} {\bibinfo {author} {\bibfnamefont {G.}~\bibnamefont {von
  Gersdorff}}\ and\ \bibinfo {author} {\bibfnamefont {W.}~\bibnamefont
  {Chen}},\ }\bibfield  {title} {\bibinfo {title} {Measurement of topological
  order based on metric-curvature correspondence},\ }\href@noop {} {\bibfield
  {journal} {\bibinfo  {journal} {Physical Review B}\ }\textbf {\bibinfo
  {volume} {104}},\ \bibinfo {pages} {195133} (\bibinfo {year}
  {2021})}\BibitemShut {NoStop}%
\bibitem [{\citenamefont {Zhang}\ \emph {et~al.}(2017)\citenamefont {Zhang},
  \citenamefont {Van Der~Laan},\ and\ \citenamefont
  {Hesjedal}}]{zhang2017direct}%
  \BibitemOpen
  \bibfield  {author} {\bibinfo {author} {\bibfnamefont {S.}~\bibnamefont
  {Zhang}}, \bibinfo {author} {\bibfnamefont {G.}~\bibnamefont {Van
  Der~Laan}},\ and\ \bibinfo {author} {\bibfnamefont {T.}~\bibnamefont
  {Hesjedal}},\ }\bibfield  {title} {\bibinfo {title} {Direct experimental
  determination of the topological winding number of skyrmions in cu2oseo3},\
  }\href@noop {} {\bibfield  {journal} {\bibinfo  {journal} {Nature
  Communications}\ }\textbf {\bibinfo {volume} {8}},\ \bibinfo {pages} {1}
  (\bibinfo {year} {2017})}\BibitemShut {NoStop}%
\bibitem [{\citenamefont {Hu}\ \emph {et~al.}(2015)\citenamefont {Hu},
  \citenamefont {Pillay}, \citenamefont {Wu}, \citenamefont {Pasek},
  \citenamefont {Shum},\ and\ \citenamefont {Chong}}]{hu2015measurement}%
  \BibitemOpen
  \bibfield  {author} {\bibinfo {author} {\bibfnamefont {W.}~\bibnamefont
  {Hu}}, \bibinfo {author} {\bibfnamefont {J.~C.}\ \bibnamefont {Pillay}},
  \bibinfo {author} {\bibfnamefont {K.}~\bibnamefont {Wu}}, \bibinfo {author}
  {\bibfnamefont {M.}~\bibnamefont {Pasek}}, \bibinfo {author} {\bibfnamefont
  {P.~P.}\ \bibnamefont {Shum}},\ and\ \bibinfo {author} {\bibfnamefont
  {Y.}~\bibnamefont {Chong}},\ }\bibfield  {title} {\bibinfo {title}
  {Measurement of a topological edge invariant in a microwave network},\
  }\href@noop {} {\bibfield  {journal} {\bibinfo  {journal} {Physical Review
  X}\ }\textbf {\bibinfo {volume} {5}},\ \bibinfo {pages} {011012} (\bibinfo
  {year} {2015})}\BibitemShut {NoStop}%
\bibitem [{\citenamefont {Hafezi}(2014)}]{hafezi2014measuring}%
  \BibitemOpen
  \bibfield  {author} {\bibinfo {author} {\bibfnamefont {M.}~\bibnamefont
  {Hafezi}},\ }\bibfield  {title} {\bibinfo {title} {Measuring topological
  invariants in photonic systems},\ }\href@noop {} {\bibfield  {journal}
  {\bibinfo  {journal} {Physical Review Letters}\ }\textbf {\bibinfo {volume}
  {112}},\ \bibinfo {pages} {210405} (\bibinfo {year} {2014})}\BibitemShut
  {NoStop}%
\bibitem [{\citenamefont {Wang}\ \emph {et~al.}(2019)\citenamefont {Wang},
  \citenamefont {Lu}, \citenamefont {Mei}, \citenamefont {Gao}, \citenamefont
  {Li}, \citenamefont {Tang}, \citenamefont {Zhu}, \citenamefont {Jia},\ and\
  \citenamefont {Jin}}]{wang2019direct}%
  \BibitemOpen
  \bibfield  {author} {\bibinfo {author} {\bibfnamefont {Y.}~\bibnamefont
  {Wang}}, \bibinfo {author} {\bibfnamefont {Y.-H.}\ \bibnamefont {Lu}},
  \bibinfo {author} {\bibfnamefont {F.}~\bibnamefont {Mei}}, \bibinfo {author}
  {\bibfnamefont {J.}~\bibnamefont {Gao}}, \bibinfo {author} {\bibfnamefont
  {Z.-M.}\ \bibnamefont {Li}}, \bibinfo {author} {\bibfnamefont
  {H.}~\bibnamefont {Tang}}, \bibinfo {author} {\bibfnamefont {S.-L.}\
  \bibnamefont {Zhu}}, \bibinfo {author} {\bibfnamefont {S.}~\bibnamefont
  {Jia}},\ and\ \bibinfo {author} {\bibfnamefont {X.-M.}\ \bibnamefont {Jin}},\
  }\bibfield  {title} {\bibinfo {title} {Direct observation of topology from
  single-photon dynamics},\ }\href@noop {} {\bibfield  {journal} {\bibinfo
  {journal} {Physical Review Letters}\ }\textbf {\bibinfo {volume} {122}},\
  \bibinfo {pages} {193903} (\bibinfo {year} {2019})}\BibitemShut {NoStop}%
\bibitem [{\citenamefont {Cardano}\ \emph {et~al.}(2017)\citenamefont
  {Cardano}, \citenamefont {D’Errico}, \citenamefont {Dauphin}, \citenamefont
  {Maffei}, \citenamefont {Piccirillo}, \citenamefont {de~Lisio}, \citenamefont
  {De~Filippis}, \citenamefont {Cataudella}, \citenamefont {Santamato},
  \citenamefont {Marrucci} \emph {et~al.}}]{cardano2017detection}%
  \BibitemOpen
  \bibfield  {author} {\bibinfo {author} {\bibfnamefont {F.}~\bibnamefont
  {Cardano}}, \bibinfo {author} {\bibfnamefont {A.}~\bibnamefont {D’Errico}},
  \bibinfo {author} {\bibfnamefont {A.}~\bibnamefont {Dauphin}}, \bibinfo
  {author} {\bibfnamefont {M.}~\bibnamefont {Maffei}}, \bibinfo {author}
  {\bibfnamefont {B.}~\bibnamefont {Piccirillo}}, \bibinfo {author}
  {\bibfnamefont {C.}~\bibnamefont {de~Lisio}}, \bibinfo {author}
  {\bibfnamefont {G.}~\bibnamefont {De~Filippis}}, \bibinfo {author}
  {\bibfnamefont {V.}~\bibnamefont {Cataudella}}, \bibinfo {author}
  {\bibfnamefont {E.}~\bibnamefont {Santamato}}, \bibinfo {author}
  {\bibfnamefont {L.}~\bibnamefont {Marrucci}}, \emph {et~al.},\ }\bibfield
  {title} {\bibinfo {title} {Detection of zak phases and topological invariants
  in a chiral quantum walk of twisted photons},\ }\href@noop {} {\bibfield
  {journal} {\bibinfo  {journal} {Nature Communications}\ }\textbf {\bibinfo
  {volume} {8}},\ \bibinfo {pages} {1} (\bibinfo {year} {2017})}\BibitemShut
  {NoStop}%
\bibitem [{\citenamefont {Grusdt}\ \emph {et~al.}(2014)\citenamefont {Grusdt},
  \citenamefont {Abanin},\ and\ \citenamefont {Demler}}]{grusdt2014measuring}%
  \BibitemOpen
  \bibfield  {author} {\bibinfo {author} {\bibfnamefont {F.}~\bibnamefont
  {Grusdt}}, \bibinfo {author} {\bibfnamefont {D.}~\bibnamefont {Abanin}},\
  and\ \bibinfo {author} {\bibfnamefont {E.}~\bibnamefont {Demler}},\
  }\bibfield  {title} {\bibinfo {title} {Measuring z 2 topological invariants
  in optical lattices using interferometry},\ }\href@noop {} {\bibfield
  {journal} {\bibinfo  {journal} {Physical Review A}\ }\textbf {\bibinfo
  {volume} {89}},\ \bibinfo {pages} {043621} (\bibinfo {year}
  {2014})}\BibitemShut {NoStop}%
\bibitem [{\citenamefont {Shen}\ and\ \citenamefont
  {Li}(2018)}]{shen2018topological}%
  \BibitemOpen
  \bibfield  {author} {\bibinfo {author} {\bibfnamefont {X.}~\bibnamefont
  {Shen}}\ and\ \bibinfo {author} {\bibfnamefont {Z.}~\bibnamefont {Li}},\
  }\bibfield  {title} {\bibinfo {title} {Topological characterization of a
  one-dimensional optical lattice with a force},\ }\href@noop {} {\bibfield
  {journal} {\bibinfo  {journal} {Physical Review A}\ }\textbf {\bibinfo
  {volume} {97}},\ \bibinfo {pages} {013608} (\bibinfo {year}
  {2018})}\BibitemShut {NoStop}%
\bibitem [{\citenamefont {Wang}\ \emph {et~al.}(2016)\citenamefont {Wang},
  \citenamefont {Xiao}, \citenamefont {Liu}, \citenamefont {Zhu},\ and\
  \citenamefont {Chan}}]{wang2016measurement}%
  \BibitemOpen
  \bibfield  {author} {\bibinfo {author} {\bibfnamefont {Q.}~\bibnamefont
  {Wang}}, \bibinfo {author} {\bibfnamefont {M.}~\bibnamefont {Xiao}}, \bibinfo
  {author} {\bibfnamefont {H.}~\bibnamefont {Liu}}, \bibinfo {author}
  {\bibfnamefont {S.}~\bibnamefont {Zhu}},\ and\ \bibinfo {author}
  {\bibfnamefont {C.~T.}\ \bibnamefont {Chan}},\ }\bibfield  {title} {\bibinfo
  {title} {Measurement of the zak phase of photonic bands through the interface
  states of a metasurface/photonic crystal},\ }\href@noop {} {\bibfield
  {journal} {\bibinfo  {journal} {Physical Review B}\ }\textbf {\bibinfo
  {volume} {93}},\ \bibinfo {pages} {041415} (\bibinfo {year}
  {2016})}\BibitemShut {NoStop}%
\bibitem [{\citenamefont {Wang}\ \emph {et~al.}(2017)\citenamefont {Wang},
  \citenamefont {Xiao}, \citenamefont {Liu}, \citenamefont {Zhu},\ and\
  \citenamefont {Chan}}]{wang2017optical}%
  \BibitemOpen
  \bibfield  {author} {\bibinfo {author} {\bibfnamefont {Q.}~\bibnamefont
  {Wang}}, \bibinfo {author} {\bibfnamefont {M.}~\bibnamefont {Xiao}}, \bibinfo
  {author} {\bibfnamefont {H.}~\bibnamefont {Liu}}, \bibinfo {author}
  {\bibfnamefont {S.}~\bibnamefont {Zhu}},\ and\ \bibinfo {author}
  {\bibfnamefont {C.~T.}\ \bibnamefont {Chan}},\ }\bibfield  {title} {\bibinfo
  {title} {Optical interface states protected by synthetic weyl points},\
  }\href@noop {} {\bibfield  {journal} {\bibinfo  {journal} {Physical Review
  X}\ }\textbf {\bibinfo {volume} {7}},\ \bibinfo {pages} {031032} (\bibinfo
  {year} {2017})}\BibitemShut {NoStop}%
\bibitem [{\citenamefont {Gao}\ \emph {et~al.}(2015)\citenamefont {Gao},
  \citenamefont {Xiao}, \citenamefont {Chan},\ and\ \citenamefont
  {Tam}}]{gao2015determination}%
  \BibitemOpen
  \bibfield  {author} {\bibinfo {author} {\bibfnamefont {W.~S.}\ \bibnamefont
  {Gao}}, \bibinfo {author} {\bibfnamefont {M.}~\bibnamefont {Xiao}}, \bibinfo
  {author} {\bibfnamefont {C.~T.}\ \bibnamefont {Chan}},\ and\ \bibinfo
  {author} {\bibfnamefont {W.~Y.}\ \bibnamefont {Tam}},\ }\bibfield  {title}
  {\bibinfo {title} {Determination of zak phase by reflection phase in 1d
  photonic crystals},\ }\href@noop {} {\bibfield  {journal} {\bibinfo
  {journal} {Optics Letters}\ }\textbf {\bibinfo {volume} {40}},\ \bibinfo
  {pages} {5259} (\bibinfo {year} {2015})}\BibitemShut {NoStop}%
\bibitem [{\citenamefont {Gupta}\ \emph
  {et~al.}(2022{\natexlab{b}})\citenamefont {Gupta}, \citenamefont {Kumar},
  \citenamefont {Tiwari}, \citenamefont {Pal}, \citenamefont {Wanare},\ and\
  \citenamefont {Ramakrishna}}]{gupta2022spectroscopic}%
  \BibitemOpen
  \bibfield  {author} {\bibinfo {author} {\bibfnamefont {N.~K.}\ \bibnamefont
  {Gupta}}, \bibinfo {author} {\bibfnamefont {M.}~\bibnamefont {Kumar}},
  \bibinfo {author} {\bibfnamefont {A.~K.}\ \bibnamefont {Tiwari}}, \bibinfo
  {author} {\bibfnamefont {S.~S.}\ \bibnamefont {Pal}}, \bibinfo {author}
  {\bibfnamefont {H.}~\bibnamefont {Wanare}},\ and\ \bibinfo {author}
  {\bibfnamefont {S.~A.}\ \bibnamefont {Ramakrishna}},\ }\bibfield  {title}
  {\bibinfo {title} {Spectroscopic ellipsometry-based investigations into the
  scattering characteristics of topologically distinct photonic stopbands},\
  }\href@noop {} {\bibfield  {journal} {\bibinfo  {journal} {Applied Physics
  Letters}\ }\textbf {\bibinfo {volume} {121}},\ \bibinfo {pages} {261103}
  (\bibinfo {year} {2022}{\natexlab{b}})}\BibitemShut {NoStop}%
\bibitem [{\citenamefont {Toll}(1956)}]{toll1956causality}%
  \BibitemOpen
  \bibfield  {author} {\bibinfo {author} {\bibfnamefont {J.~S.}\ \bibnamefont
  {Toll}},\ }\bibfield  {title} {\bibinfo {title} {Causality and the dispersion
  relation: logical foundations},\ }\href@noop {} {\bibfield  {journal}
  {\bibinfo  {journal} {Physical review}\ }\textbf {\bibinfo {volume} {104}},\
  \bibinfo {pages} {1760} (\bibinfo {year} {1956})}\BibitemShut {NoStop}%
\bibitem [{\citenamefont {Ramakrishna}\ and\ \citenamefont
  {Grzegorczyk}(2008)}]{ramakrishna2008physics}%
  \BibitemOpen
  \bibfield  {author} {\bibinfo {author} {\bibfnamefont {S.~A.}\ \bibnamefont
  {Ramakrishna}}\ and\ \bibinfo {author} {\bibfnamefont {T.~M.}\ \bibnamefont
  {Grzegorczyk}},\ }\href@noop {} {\emph {\bibinfo {title} {Physics and
  applications of negative refractive index materials}}}\ (\bibinfo
  {publisher} {CRC press},\ \bibinfo {year} {2008})\BibitemShut {NoStop}%
\bibitem [{\citenamefont {Venema}\ and\ \citenamefont
  {Schmidt}(2008)}]{venema2008optical}%
  \BibitemOpen
  \bibfield  {author} {\bibinfo {author} {\bibfnamefont {T.~M.}\ \bibnamefont
  {Venema}}\ and\ \bibinfo {author} {\bibfnamefont {J.~D.}\ \bibnamefont
  {Schmidt}},\ }\bibfield  {title} {\bibinfo {title} {Optical phase unwrapping
  in the presence of branch points},\ }\href@noop {} {\bibfield  {journal}
  {\bibinfo  {journal} {Optics Express}\ }\textbf {\bibinfo {volume} {16}},\
  \bibinfo {pages} {6985} (\bibinfo {year} {2008})}\BibitemShut {NoStop}%
\bibitem [{\citenamefont {Asb{\'o}th}\ \emph {et~al.}(2016)\citenamefont
  {Asb{\'o}th}, \citenamefont {Oroszl{\'a}ny},\ and\ \citenamefont
  {P{\'a}lyi}}]{asboth2016short}%
  \BibitemOpen
  \bibfield  {author} {\bibinfo {author} {\bibfnamefont {J.~K.}\ \bibnamefont
  {Asb{\'o}th}}, \bibinfo {author} {\bibfnamefont {L.}~\bibnamefont
  {Oroszl{\'a}ny}},\ and\ \bibinfo {author} {\bibfnamefont {A.}~\bibnamefont
  {P{\'a}lyi}},\ }\bibfield  {title} {\bibinfo {title} {A short course on
  topological insulators},\ }\href@noop {} {\bibfield  {journal} {\bibinfo
  {journal} {Lecture notes in physics}\ }\textbf {\bibinfo {volume} {919}},\
  \bibinfo {pages} {166} (\bibinfo {year} {2016})}\BibitemShut {NoStop}%
\bibitem [{\citenamefont {Tan}\ \emph {et~al.}(2014)\citenamefont {Tan},
  \citenamefont {Sun}, \citenamefont {Chen},\ and\ \citenamefont
  {Shen}}]{tan2014photonic}%
  \BibitemOpen
  \bibfield  {author} {\bibinfo {author} {\bibfnamefont {W.}~\bibnamefont
  {Tan}}, \bibinfo {author} {\bibfnamefont {Y.}~\bibnamefont {Sun}}, \bibinfo
  {author} {\bibfnamefont {H.}~\bibnamefont {Chen}},\ and\ \bibinfo {author}
  {\bibfnamefont {S.-Q.}\ \bibnamefont {Shen}},\ }\bibfield  {title} {\bibinfo
  {title} {Photonic simulation of topological excitations in metamaterials},\
  }\href@noop {} {\bibfield  {journal} {\bibinfo  {journal} {Scientific
  Reports}\ }\textbf {\bibinfo {volume} {4}},\ \bibinfo {pages} {1} (\bibinfo
  {year} {2014})}\BibitemShut {NoStop}%
\bibitem [{\citenamefont {Huang}\ \emph {et~al.}(2019)\citenamefont {Huang},
  \citenamefont {Guo}, \citenamefont {Feng}, \citenamefont {Yu}, \citenamefont
  {Jiang}, \citenamefont {Zhang}, \citenamefont {Wang},\ and\ \citenamefont
  {Chen}}]{huang2019observation}%
  \BibitemOpen
  \bibfield  {author} {\bibinfo {author} {\bibfnamefont {Q.}~\bibnamefont
  {Huang}}, \bibinfo {author} {\bibfnamefont {Z.}~\bibnamefont {Guo}}, \bibinfo
  {author} {\bibfnamefont {J.}~\bibnamefont {Feng}}, \bibinfo {author}
  {\bibfnamefont {C.}~\bibnamefont {Yu}}, \bibinfo {author} {\bibfnamefont
  {H.}~\bibnamefont {Jiang}}, \bibinfo {author} {\bibfnamefont
  {Z.}~\bibnamefont {Zhang}}, \bibinfo {author} {\bibfnamefont
  {Z.}~\bibnamefont {Wang}},\ and\ \bibinfo {author} {\bibfnamefont
  {H.}~\bibnamefont {Chen}},\ }\bibfield  {title} {\bibinfo {title}
  {Observation of a topological edge state in the x-ray band},\ }\href@noop {}
  {\bibfield  {journal} {\bibinfo  {journal} {Laser \& Photonics Reviews}\
  }\textbf {\bibinfo {volume} {13}},\ \bibinfo {pages} {1800339} (\bibinfo
  {year} {2019})}\BibitemShut {NoStop}%
\bibitem [{\citenamefont {Shi}\ \emph {et~al.}(2016)\citenamefont {Shi},
  \citenamefont {Xue}, \citenamefont {Jiang},\ and\ \citenamefont
  {Chen}}]{shi2016topological}%
  \BibitemOpen
  \bibfield  {author} {\bibinfo {author} {\bibfnamefont {X.}~\bibnamefont
  {Shi}}, \bibinfo {author} {\bibfnamefont {C.}~\bibnamefont {Xue}}, \bibinfo
  {author} {\bibfnamefont {H.}~\bibnamefont {Jiang}},\ and\ \bibinfo {author}
  {\bibfnamefont {H.}~\bibnamefont {Chen}},\ }\bibfield  {title} {\bibinfo
  {title} {Topological description for gaps of one-dimensional symmetric
  all-dielectric photonic crystals},\ }\href@noop {} {\bibfield  {journal}
  {\bibinfo  {journal} {Optics Express}\ }\textbf {\bibinfo {volume} {24}},\
  \bibinfo {pages} {18580} (\bibinfo {year} {2016})}\BibitemShut {NoStop}%
\bibitem [{\citenamefont {Gupta}\ \emph
  {et~al.}(2022{\natexlab{c}})\citenamefont {Gupta}, \citenamefont {Singh},
  \citenamefont {Srinivasu}, \citenamefont {Wanare}, \citenamefont
  {Srivastava}, \citenamefont {Ramkumar},\ and\ \citenamefont
  {Ramakrishna}}]{gupta2022singular}%
  \BibitemOpen
  \bibfield  {author} {\bibinfo {author} {\bibfnamefont {N.~K.}\ \bibnamefont
  {Gupta}}, \bibinfo {author} {\bibfnamefont {G.}~\bibnamefont {Singh}},
  \bibinfo {author} {\bibfnamefont {S.}~\bibnamefont {Srinivasu}}, \bibinfo
  {author} {\bibfnamefont {H.}~\bibnamefont {Wanare}}, \bibinfo {author}
  {\bibfnamefont {K.~V.}\ \bibnamefont {Srivastava}}, \bibinfo {author}
  {\bibfnamefont {J.}~\bibnamefont {Ramkumar}},\ and\ \bibinfo {author}
  {\bibfnamefont {S.~A.}\ \bibnamefont {Ramakrishna}},\ }\bibfield  {title}
  {\bibinfo {title} {Singular-phase characteristics of electromagnetic
  absorbers and a framework for low-rcs target detection},\ }\href@noop {}
  {\bibfield  {journal} {\bibinfo  {journal} {IEEE Antennas and Wireless
  Propagation Letters}\ } (\bibinfo {year} {2022}{\natexlab{c}})}\BibitemShut
  {NoStop}%
\end{thebibliography}%

\end{document}